\documentclass[aps,prd,nofootinbib,showpacs,superscriptaddress]{revtex4} 
\usepackage[sumlimits]{amsmath}
\usepackage{epsfig}
\usepackage{amssymb}
\usepackage{graphicx}
\usepackage{pstricks}
\usepackage{color}
\usepackage{bbold}
\usepackage{xfrac}
\usepackage{hyperref}

\newcommand{\dd}{\mathrm{d}}

\begin{document}

%
%

\title{Inflation from a chaotic potential with a step}
\author{Clara Rojas} 
\email{crojas@yachaytech.edu.ec}
\affiliation{Yachay Tech University, School of Physical Sciences and Nanotechnology, Hda. San Jos\'{e} s/n y \\
Proyecto Yachay, 100119, Urcuqu\'{i}, Ecuador}
\author{Rafael Hern\'{a}ndez-Jim\'{e}nez}
\email{rafaelhernandezjmz@gmail.com}
\affiliation{Departamento de F\'isica,
Centro Universitario de Ciencias Exactas e Ingenier\'ia, Universidad de Guadalajara\\
Av. Revoluci\'on 1500, Colonia Ol\'impica C.P. 44430, Guadalajara, Jalisco, M\'exico}
%
%
\date{\today}

%
%
\bigskip
\begin{abstract}
In this work, we study the effects on the relevant observational parameters of an inflationary universe from a chaotic potential with a step. We numerically evolve the perturbation equations within both cold inflation and warm inflation. On the one hand, in a cold inflation scenario we analyse the scalar power spectrum $P_{\mathcal{R}}$ in terms of the number of e-folds $N_{e}$, and in terms of the ratio $k/k_{0}$, where $k_{0}$ is our pivot scale. We show how $P_{\mathcal{R}}$ oscillates around $0.2< k/k_{0} < 20$. Additionally, we present the evolution of two relevant parameters: the scalar spectral index $n_\mathrm{s}$ and the tensor-to-scalar ratio $r$. In fact, more than one region of $(n_\mathrm{s},r)$ lies within the observable window (Planck 2018). On the other hand, in the warm inflationary case, we also examine the evolution of $P_{\mathcal{R}}$ in terms of $N_{e}$ and $k/k_{0}$. Perturbations are amplified in WI; in fact, $P_{\mathcal{R}}$ can be much larger than the CMB value $P_{\mathcal{R}}> 2.22\times 10^{-9}$. This time, the spectral index $n_\mathrm{s}$ is clearly blue-tilted, at smaller scales, and the tensor-to-scalar ratio $r$ becomes too low. However, $n_\mathrm{s}$ can change from blue-tilted towards red-tilted, since $P_{\mathcal{R}}$ starts oscillating around $k/k_{0}\sim 40$. Indeed, the result from the step potential skims the Planck contours. Finally, one key aspect of this research was to contrast the features of an inflationary potential between both paradigms, and, in fact, they show similarities and differences. Due to a featured background and a combined effect of entropy fluctuations (only in warm inflation), in both scenarios certain fluctuation scales are not longer ``freeze in'' on super-horizon scales. In contrast, warm inflation boosts the growth of the amplitude of the primordial spectrum, whilst moving the oscillatory behaviour from the feature scale $k/k_{0}\sim 1$ (manifested in cold inflation) to $k/k_{0}\sim 40$; although not only the step potential wiggles, but also the chaotic. This might indicate that warm inflation could function as a screening mechanism capable of smoothing the feature of the inflationary potential; however, further investigation is required to corroborate this. 
%
\end{abstract}



\maketitle
%
%
\section{Introduction}
The Standard Big Bang (SBB) paradigm successfully describes the evolution of the universe at large scales. This theory rests upon four fundamental pillars, a theoretical framework based on General Relativity, as put forward by Einstein and Friedman, and three robust observational facts: $1$) the expansion of the universe observed by Hubble; $2$) the observed abundance of light elements predicted by Gamow and his collaborators; and $3$) the presence of the Cosmic Microwave Background (CMB) of a very isotropic black-body radiation at a temperature of about $3\, \mathrm{K}$, which presents a high degree of uniformity with inhomogeneities of about one part in $10^5$ \cite{bellido:2000}. Due to the observation of the CMB by Penzias and Wilson, the SBB theory became the main candidate to describe our universe and therefore provides a framework for studying its history from $10^{-2}\,\mathrm{s}$ to our present age, around $13.7\, \mathrm{Gyr}$. However, this cosmological model yields several shortcomings. Two of the most relevant deficiencies are flatness and horizon problems.

Guth realised that a period of very rapid expansion at early times (that happened between $10^{-34\pm 6}\,\mathrm{s}$ and $10^{-32\pm 6} \,\mathrm{s}$ \cite{liddle:2000}), called the inflationary universe, solved the flaws of the SBB theory \cite{guth:1981}. Indeed, inflation predicts a flat universe where all relevant scales were casually connected; thus, the SBB framework does not have these issues. The inflationary paradigm is mainly characterised by the existence of a homogeneous scalar field, the inflaton $\phi$, which is the cause of an abrupt exponential expansion, from a specific inflaton potential, producing different scenarios. It is also characterised by the presumed presence of quantum fluctuations $\delta\phi$ around $\phi$. It is this last feature that provides a clever mechanism to generate a primordial spectrum of density perturbations \cite{chibisov:1982} (or primordial power spectrum) almost scale invariant (a power-law) and nearly Gaussian, which might explain the anisotropies of the CMB temperature map, hence generating the seeds for large-scale structure in the universe.

In the standard inflationary picture, called cold inflation (CI), the super-fast period of accelerated expansion quickly dilutes away all traces of any pre-inflationary matter or radiation density, leaving this framework in a vacuum state. However, this yields a supercooled universe. Therefore, to successfully explain the transition from inflation to the SBB scenario and hence the physics of recombination leading to the CMB that we observe today, the inflaton energy density necessarily requires that it be transferred into ordinary matter and radiation, and thus to its interactions with other fields. Accordingly, the inflaton field could be coupled to other components and might dissipate its vacuum energy and warm up the universe. This alternative scenario is known as the warm inflation (WI) paradigm \cite{Berera:1995wh,Berera:1995ie,Berera:1996fm,Berera:2008ar}, where dissipative effects and associated particle production can, in fact, sustain a thermal bath concurrently with the accelerated expansion of the universe during inflation. Therefore, this enhanced mechanism becomes more relevant in phenomenology. Certainly, WI has attractive characteristics. For example, even if radiation is subdominant during inflation, it can easily become the leading component at the end of inflation, without the need for a separate reheating or preheating period. Moreover, dissipation also affects scalar perturbations, albeit for WI the fluctuations of the inflaton are thermally induced, which are already classical upon definition; hence, they may bring an interpretation of the nature of the classical inhomogeneities observed in the CMB; consequently, there is no need to explain the troublesome quantum-to-classical transition problem of CI, due to the purely quantum origin of the CI density perturbations. Importantly, many scenarios of WI have been observed to be within the observable window~\cite{Bastero-Gil:2016qru, Bastero-Gil:2018uep, Bastero-Gil:2018yen}. Moreover, it has been shown that WI can avoid the proposed swampland conjectures~\cite{Motaharfar:2018zyb, Das:2018rpg, Bastero-Gil:2018yen}. Thus, WI is in accordance with both the current cosmological observations and the proposed Swampland Criteria. 
	
\begin{figure}[htbp] 
\includegraphics[scale=0.7]{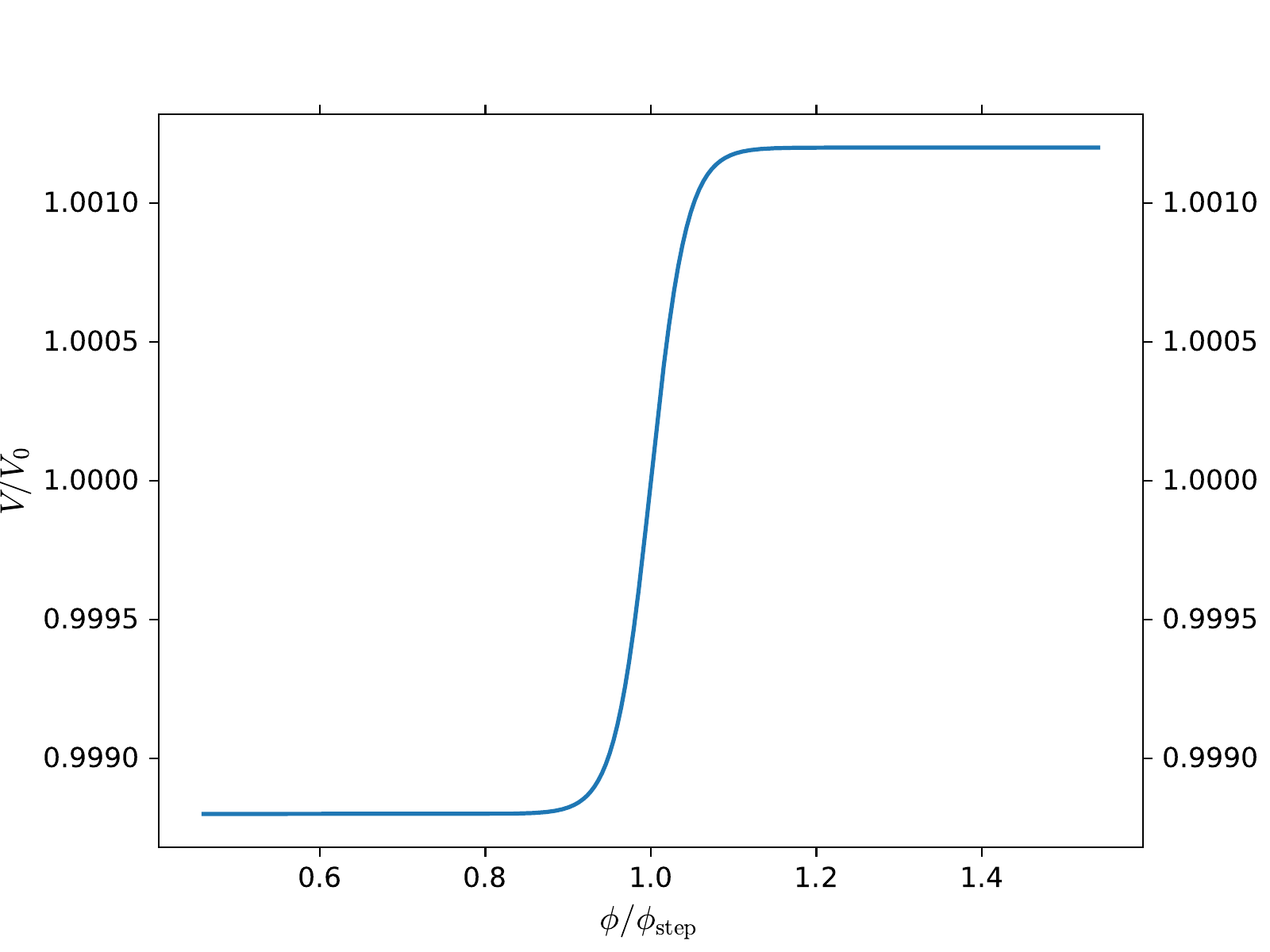}
\caption{Evolution of $V/V_{0}$ as a function of the normalised inflaton field $\phi/\phi_{\textnormal{step}}$. We take $m=5.731\times 10^{-6}\, M_{Pl}$, $\phi_{\textnormal{step}}=15.451 M_{Pl}$, $c=0.0012$, $d=0.04  \, M_{Pl}$, and the initial value of the inflaton $\phi_{0}=16.706 \, M_{Pl}$, where $M_{Pl} = 1/\sqrt{8\pi G}$ is the reduced Planck mass. Note that the step occurs at $\phi/\phi_{\textnormal{step}}\simeq 1$.}\label{fig:step_potential}
\end{figure}

On the other hand, the Planck 2018 results favour inflation \cite{akrami:2018a}. Indeed, the data is consistent with a flat universe, almost scale invariant, and nearly Gaussian primordial power spectrum. Different inflation models lead to distinct predictions; however, one has to be able to discriminate among existing templates using the recent CMB data and, in addition, to determine the shape of the primordial power spectrum. Although Planck $2018$ results do not provide statistically significant evidence for features in the angular power spectrum~\cite{akrami:2018a}, this possibility has been explored and widely discussed, first by Starorbinsky~\cite{starobinsky:1992}, and later by a number of authors~\cite{adams:2001,hunt:2004,covi:2006,hamann:2007,chen:2007,hunt:2007,hamann:2010,bartolo:2013,cadavid:2015,cadavid:2017}. These features could be caused by a step in the inflationary potential that leads to oscillations in the scalar power spectrum. Adams~\cite{adams:2001} has proposed a particular model in which a step feature is added to the chaotic inflationary potential, $V_{0}=m^2\phi^2/2$, in the following way (see Fig.~\ref{fig:step_potential}):
\begin{equation}
\label{potential}
V(\phi) = \dfrac{1}{2}m^2\phi^{2}\left[1+c\,\tanh\left(\frac{\phi-\phi_\textnormal{step}}{d}\right)\right],   
\end{equation}
where the step occurs at $\phi=\phi_\textnormal{step}$, $m$ is the inflaton mass, and the parameters $c$ and $d$ are related to the amplitude and width of the feature, respectively. 
The presence of the step does not interrupt inflation, but introduces oscillations in the primordial power spectrum that depend on the height and gradient of the step, and as a consequence it will be scale dependent.

In this work, we present an analysis of an inflationary framework of a chaotic potential with a step from the CI and WI perspectives. We numerically solve all background and linear-perturbed variables, which in turn will allow us to scan all relevant parameters along a large range of scales. We take a pivot wave number $k_{0}$ when the step occurs. On the one hand, we thoroughly study the CI case, where we find outcome regions within the observational window. On the other hand, the WI instance is presented, where we show the main differences between the CI and WI scenarios. 

This paper is organised as follows. On the one hand, in Section II we exhaustively study a CI scenario. On the other hand, in Section III we describe a WI case. Section IV shows a comparative study between CI and WI. Finally, in Section V we give the conclusions and outlook of this work.
%
\section{Single field inflation: Cold Inflation}
%
We start with a model described by the non-zero vacuum expectation value of the inflaton field $\phi(t)$ in the homogeneous and isotropic state; this carries most of the energy of the universe, and hence any other matter content must be subdominant. Additionally, its fluctuation $\delta\phi(\mathbf{x},t)$, which describes the quantum fluctuation around $\phi$. We will work within a homogeneous and isotropic flat Friedmann-Lema\^{i}tre-Robertson-Walker (FLRW) metric: 
\begin{equation}\label{RW1}
\dd s^{2}= -\dd t^{2}+a(t)^{2}\delta_{ij}\dd x^{i}\dd x^{j}\,,
\end{equation}
where $t$ is the cosmological time, $a=a(t)$ is the scale factor. Then, in the cosmological case, the universe inflates as the field is rolling down the hill. The dynamic equations are as follows 
\begin{eqnarray} 
&& H^{2} = \frac{\rho_{\phi}}{3M_{Pl}^{2}} \,,\label{CI-equations1}\\
&& \ddot{\phi}+3H\dot{\phi}+V_{\phi} = 0 \,, \label{CI-equations2}
\end{eqnarray}
where dots represent derivatives with respect to time $t$; then $\rho_{\phi}=\dot{\phi}^{2}/2+V(\phi)$ is the energy density of the inflaton, $V_{\phi}$ is the derivative of the potential energy with respect to the field, and $H=\dot{a}/a$ is the expansion rate or Hubble parameter. In addition, $M_{Pl} = 1/\sqrt{8\pi G}$ is the reduced Planck mass. The first expression is called Friedmann constraint, and the second one is the scalar field equation Klein-Gordon or the energy-momentum conservation equation. As we mentioned before, inflation requires, for example, that $\rho_{r}$ (radiation)$\ll \rho_{\phi}$ to generate an abrupt expansion; however, any other subdominant component present could play an important role at the end of inflation. Indeed, this is the study case for the next section of this paper. 
The amount by which the universe inflates is measured as the number of e-foldings $N_{e}$, and since the size of the expansion is expected to be an enormous quantity, it is useful to compute it in terms of the logarithm of the ratio of the scale factor between the end of inflation and a time $t$ during inflation defined by:
\begin{equation}\label{Ne-phi}
N_{e} \equiv \ln\left[\frac{a(t_{end})}{a(t)}\right]=\int_{t}^{t_{end}}\dd t\, H \,.
\end{equation}
Typically between $40$ and $60$ e-foldings of observable inflation~\footnote{CMB fluctuations, from the size of the observable universe down to the size of galaxies, were generated during $\sim$10 e-folds about $60$ e-folds before the end of inflation.} are large enough to solve the horizon and flatness problems. Finally, inflation lasts while the slow-roll parameter $\epsilon_{H}=-\dot{H}/H^{2}<1$, where $\dot{H} = -\dot{\phi}^{2}/(2M_{Pl}^{2})$, so it ends when $\epsilon_{H}=1$. An inflationary scenario also requires a flat potential, and this condition is measured by the $\eta_{H}=-\ddot{\phi}/(H\dot{\phi})$ parameter, where a particular potential preserves the flatness having $|\eta_{H}|<1$. 
For the time being, above system, Eqs. (\ref{CI-equations1},\ref{CI-equations2}),  represents only the background dynamics. However, we must introduce the fluctuations of such a physical system. First, we present the spacetime metric for a scalar-type of perturbation in the \emph{zero-shear gauge} is given by \cite{Hwang:1991aj}:
\begin{equation}\label{perturbed-metric}
\dd s^{2}=-(1+2\alpha)\dd t^{2}+a^{2}\left[\delta_{ij}(1-2\alpha)\right]\dd x^{i}\dd x^{j} \,,
\end{equation}
where $\alpha$ is a spacetime-dependent perturbed-order variable. Then, for the inflaton, we replace the scalar field by $\phi + \delta\phi$, and using the perturbed metric Eq. \eqref{perturbed-metric} working in momentum space, that is, in Fourier space, we get the first order perturbation equations \cite{Hwang:2001fb, Kodama:1984ziu}: 
\begin{eqnarray}
&& \ddot{\delta\phi} + 3H\dot{\delta\phi} + \left(\frac{k^{2}}{a^{2}}+V_{\phi\phi}  \right)\delta\phi = 4\dot{\alpha}\dot{\phi} + (2\ddot{\phi}+6H\dot{\phi})\alpha  \,,\\
&& \ddot{\alpha} + 4H\dot{\alpha} + (3H^{2}+2\dot{H})\alpha = \frac{\dot{\phi}\dot{\delta\phi}-\dot{\phi}^{2}\alpha-V_{\phi}\delta\phi}{2M_{Pl}^{2}} \,,
\end{eqnarray}
where $k$ is the comoving wavenumber. We will evolve the equations for the perturbations at linear order, for a given model, without approximations and working in the \emph{zero-shear gauge}. For a scalar field, the power spectrum of the comoving curvature perturbation $\mathcal{R}$ is as follows:
\begin{equation}
P_{\mathcal{R}} = \frac{k^{3}}{2\pi^{2}}\left(\frac{H}{\dot{\phi}}\right)^{2}\left|\delta\phi^{GI}\right|^{2} \,,\qquad {\rm where} \, \qquad  \delta\phi^{GI} = \delta\phi + \frac{\dot{\phi}}{H}\alpha \,.
\end{equation}
To contrast the observable parameters with Planck data, we need to evaluate, at horizon crossing, the scalar spectral index $n_\mathrm{s}$ and the tensor-to-scalar ratio $r$ which are defined as  
\begin{equation}
n_\mathrm{s}-1\equiv \frac{d\ln P_{\mathcal{R}}}{d\ln k} \,,\quad r=\frac{P_{h}}{P_{\mathcal{R}}} \,,
\end{equation}
where $P_{h}$ is the tensor power spectrum {\footnote{Tensor perturbations are described by: $\ddot{h}_{k}+3H\dot{h}_{k}+\frac{k^{2}}{a^{2}}h_{k} = 0$. And the tensor power spectrum is $P_{h} = \frac{4 k^{3}}{\pi^{2}}\left|h_{k}\right|^{2}$.}}. However, numerically, we evaluate $n_\mathrm{s}$ and $r$ when $P_{\mathcal{R}}$ and $P_{h}$ become constant, approximately twice after horizon crossing. 

\subsection{CI results:}
To study the effects of features on the spectrum of primordial perturbations, we consider the following numerical example. We take $m=5.731\times 10^{-6}\, M_{Pl}$, $\phi_{\textnormal{step}}=15.451 M_{Pl}$, $c=0.0012$, $d=0.04  \, M_{Pl}$, and the initial value of the inflaton $\phi_{0}=16.706 \, M_{Pl}$. Inflation lasts $N_{e}=70.2$, and the step occurs at $\phi=\phi_{\textnormal{step}}$ around $N_{e}\simeq 10.1$. Therefore, from the time of the step about $60$ e-folds of inflation take place. 
\begin{figure}[htbp] 
\includegraphics[scale=0.548]{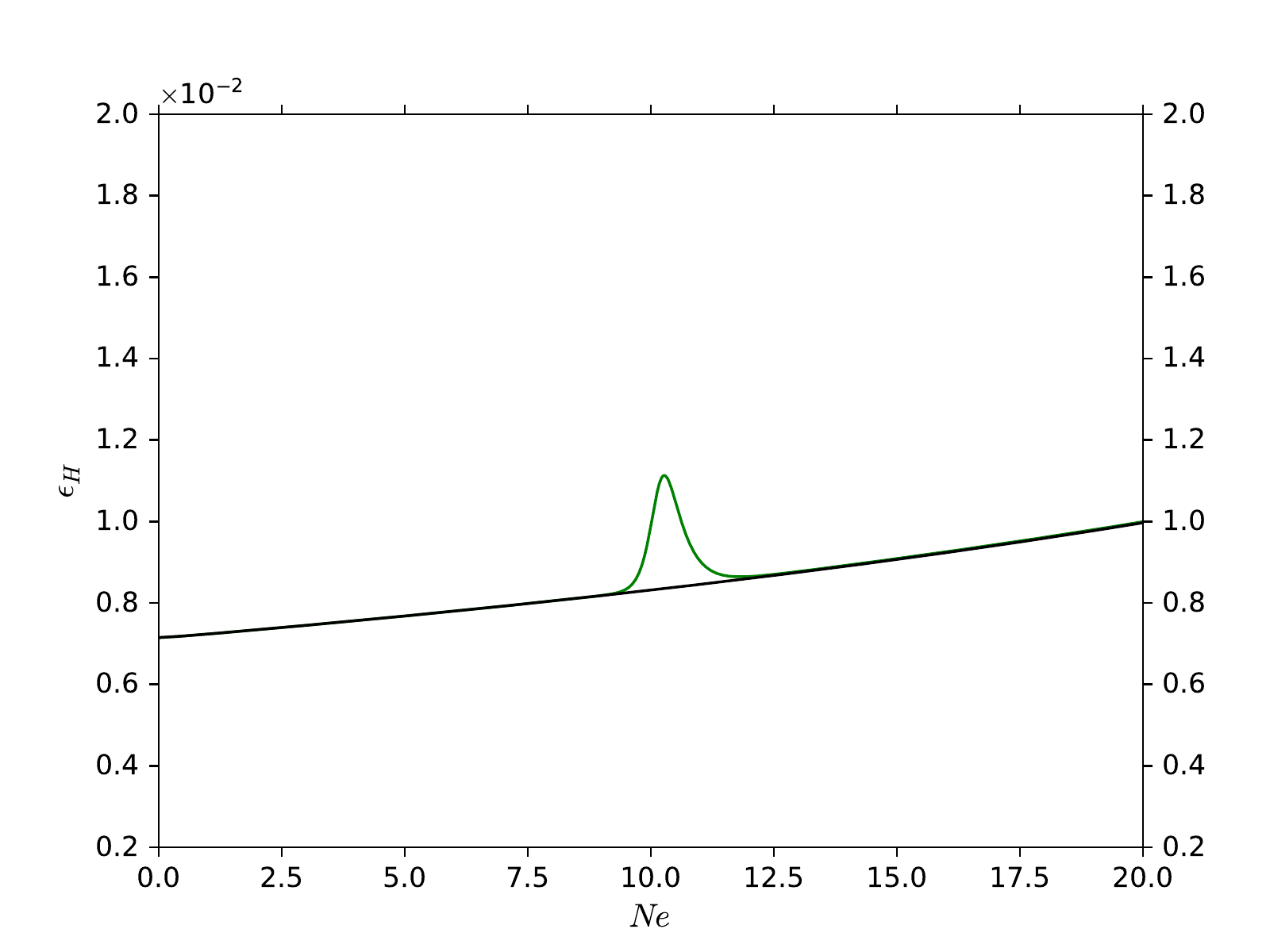}
\includegraphics[scale=0.548]{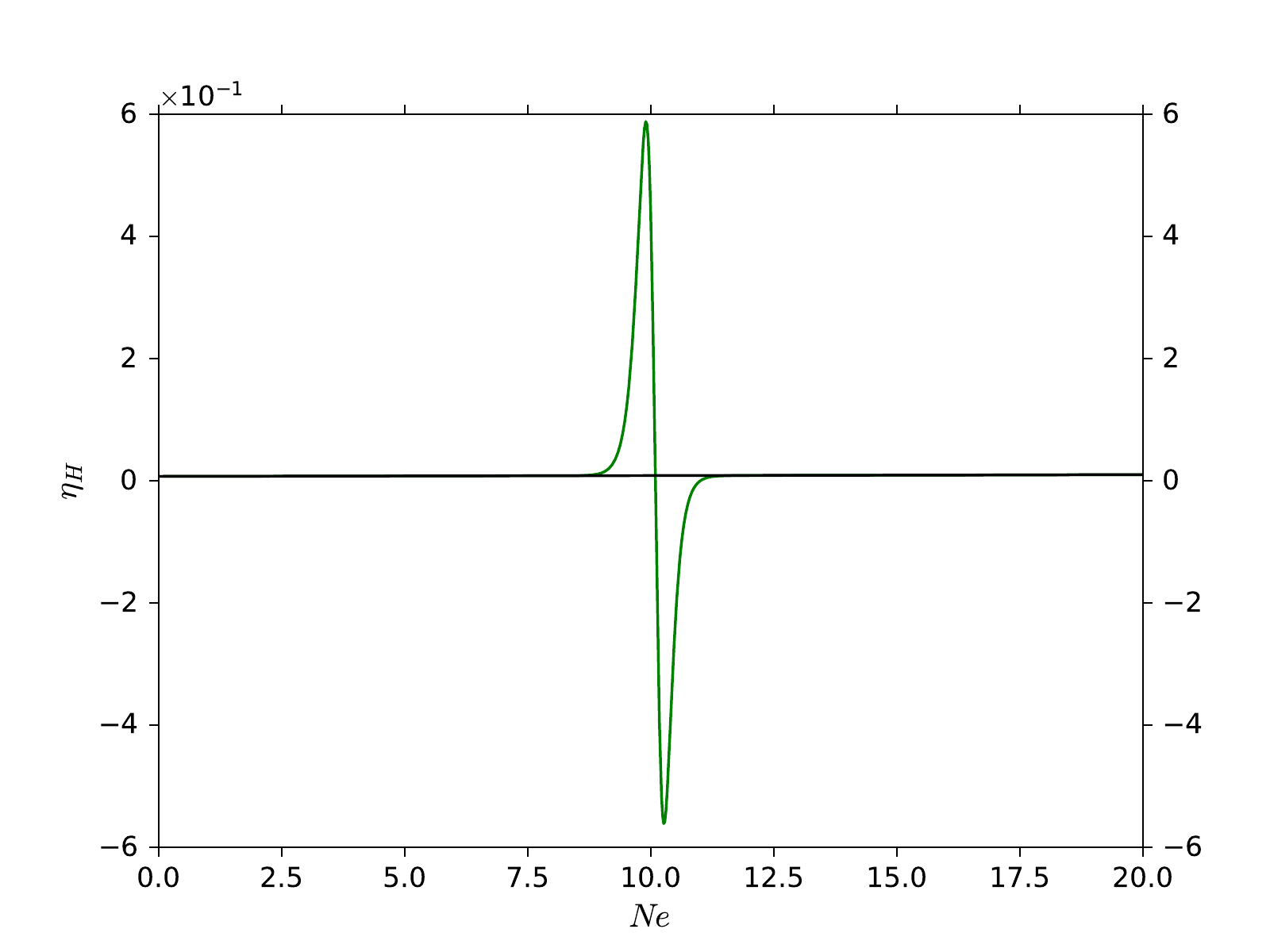} 
\caption{Behaviour of the slow-roll parameters $\epsilon_{H}$ and $\eta_{H}$ with respect to the number of e-folds $N_{e}$ within the CI scheme, described by step (green) and quadratic chaotic (black) potentials for $70.2$ e-folds of inflation. Note that the step occurs at $N_{e}\simeq 10$.}\label{fig:CI_slow_roll_parameters}
\end{figure}

In Fig.~\ref{fig:CI_slow_roll_parameters} we show the effects of the features on the slow-roll parameters with respect to the number of e-foldings $N_{e}$. The black line characterises a quadratic chaotic description, whilst the green line corresponds to a step potential. Indeed, the feature occurs around $N_{e}\simeq 10$, from where both schemes evolve equally. This upshot will be determined when analysing the linear perturbations. 

Before continuing our analysis, we must explicitly explain how to calculate the pivot scale $k_{0}$. Since we numerically solve all variables, we estimate the feature of the potential when $\phi/\phi_{\mathrm{step}} \simeq 1$ at a particular $N_{e}(\mathrm{step})$, from which we equate $k_{0}=a(N_{e}(\mathrm{step}))H(N_{e}(\mathrm{step}))$. Thus, we set our reference value $k_{0}$ when the step occurs.   

\begin{figure}[htbp] 
\includegraphics[scale=0.90]{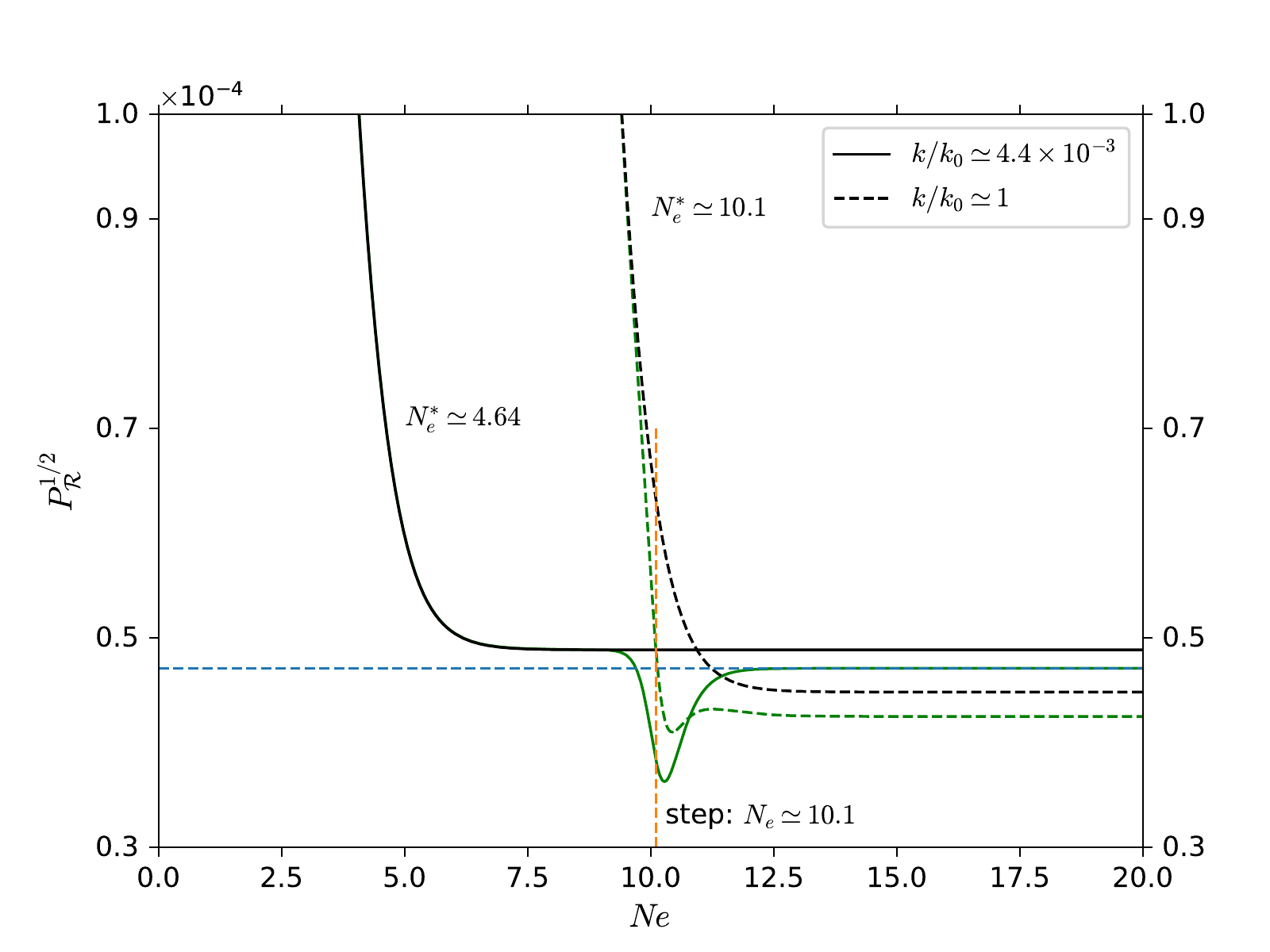}
\caption{Behaviour of the square root of the curvature power spectra $P_{\mathcal{R}}^{1/2}$ with respect to the number of e-folds $N_{e}$, described by step (green) and quadratic chaotic (black) potentials for $70.2$ e-folds of inflation. We show the outcome for two different wave numbers $k/k_{0} \simeq 1, 4.4\times 10^{-3}$, where $k_{0}\simeq 0.0256 \,h\,\rm Mpc^{-1}$. Solid lines correspond to $N_{e}^{*}\simeq 4.64$, whilst dashed lines correspond to $N_{e}^{*}\simeq 10.1$. The blue dashed line represents the CMB observations by the Planck Legacy value $P_{\mathcal{R}}^{1/2}\simeq 4.7\times 10^{-5}$ \cite{akrami:2018b}.}\label{fig:CI_PR}
\end{figure}

Then, Fig.~\ref{fig:CI_PR} shows the behaviour of the square root of the curvature power spectra $P_{\mathcal{R}}^{1/2}$ with respect to $N_{e}$. We only present the outcome of two distinct scales $k/k_{0} \simeq 1, 4.4\times 10^{-3}$, where $k_{0}\simeq 0.0256 \,h\rm\,Mpc^{-1}$ {\footnote{The value of $k_{0}$ is evaluated in physical units. We follow Salman Habib, et al.~\cite{Habib:2005mh}, conversion to physical units. From Appendix 3, the rescaled momentum $k$ in physical units $h\rm\,Mpc^{-1}$ is given by
\begin{equation}
k_{phys} = \sqrt{8\pi}\frac{h}{100 c a(0)} k \, \rm km\,s^{-1}\,Mpc^{-1} \,,
\end{equation}
where $c=2.99792458\times 10^{5}\rm km\,s^{-1}$ is the speed of light, the initial expansion rate is $a(0)=1$ (in our code), and $k$ is the dimensionless momentum used in the code. Throughout the paper, we do not use the suffix ``phys''; however, all results are given in physical units.}}, and $h=H_{0}/100 = 0.6732$, see \cite{akrami:2018b, Planck:2018jri}. This figure helps us to illustrate how the step spectrum behaves compared with the featureless one. Actually, many of our results are very similar to \cite{adams:2001,hunt:2004,covi:2006,hamann:2007,chen:2007,hunt:2007,hamann:2010,bartolo:2013,cadavid:2015,cadavid:2017}, and in particular to \cite{cadavid:2017}. Moreover, note that even the solid-lined case, when horizon crossing $N_{e}^{*}\simeq 4.64$, $P_{\mathcal{R}}^{1/2}$ abruptly changes when the step occurs, which might indicate that this particular perturbation is not longer ``freeze in'' on super-horizon scales; however, after a short period this featured comoving curvature fluctuation becomes, in fact, constant; therefore, its horizon reentry value is slightly different from the horizon exit one. 

\begin{figure}[htbp] 
\includegraphics[scale=0.90]{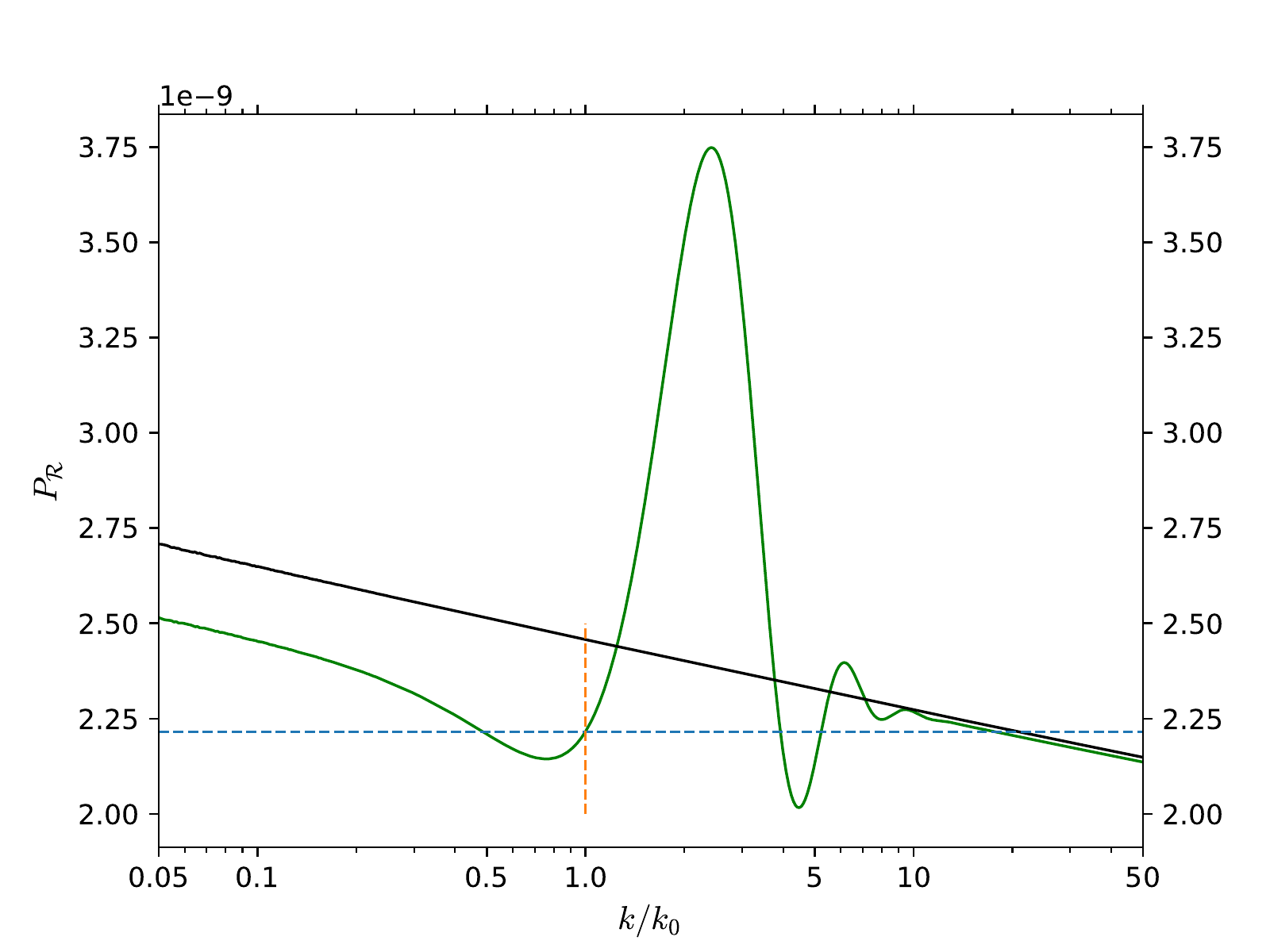}
\caption{Behaviour of the curvature power spectra $P_{\mathcal{R}}$ with respect to the ratio $k/k_{0}$, where $k_{0}\simeq 0.04  \,h \,\rm Mpc^{-1}$, described by step (green) and quadratic chaotic (black) potentials for $71.0$ e-folds of inflation. The orange vertical line corresponds to $k/k_{0}=1$ and $P_{\mathcal{R}}\simeq 2.22\times 10^{-9}$. On the other hand, the blue dashed line corresponds to the CMB observations by the Planck Legacy value $P_{\mathcal{R}}\simeq 2.22\times 10^{-9}$ \cite{akrami:2018b}.}\label{fig:CI_PR_k}
\end{figure}

\begin{figure}[htbp] 
\includegraphics[scale=0.548]{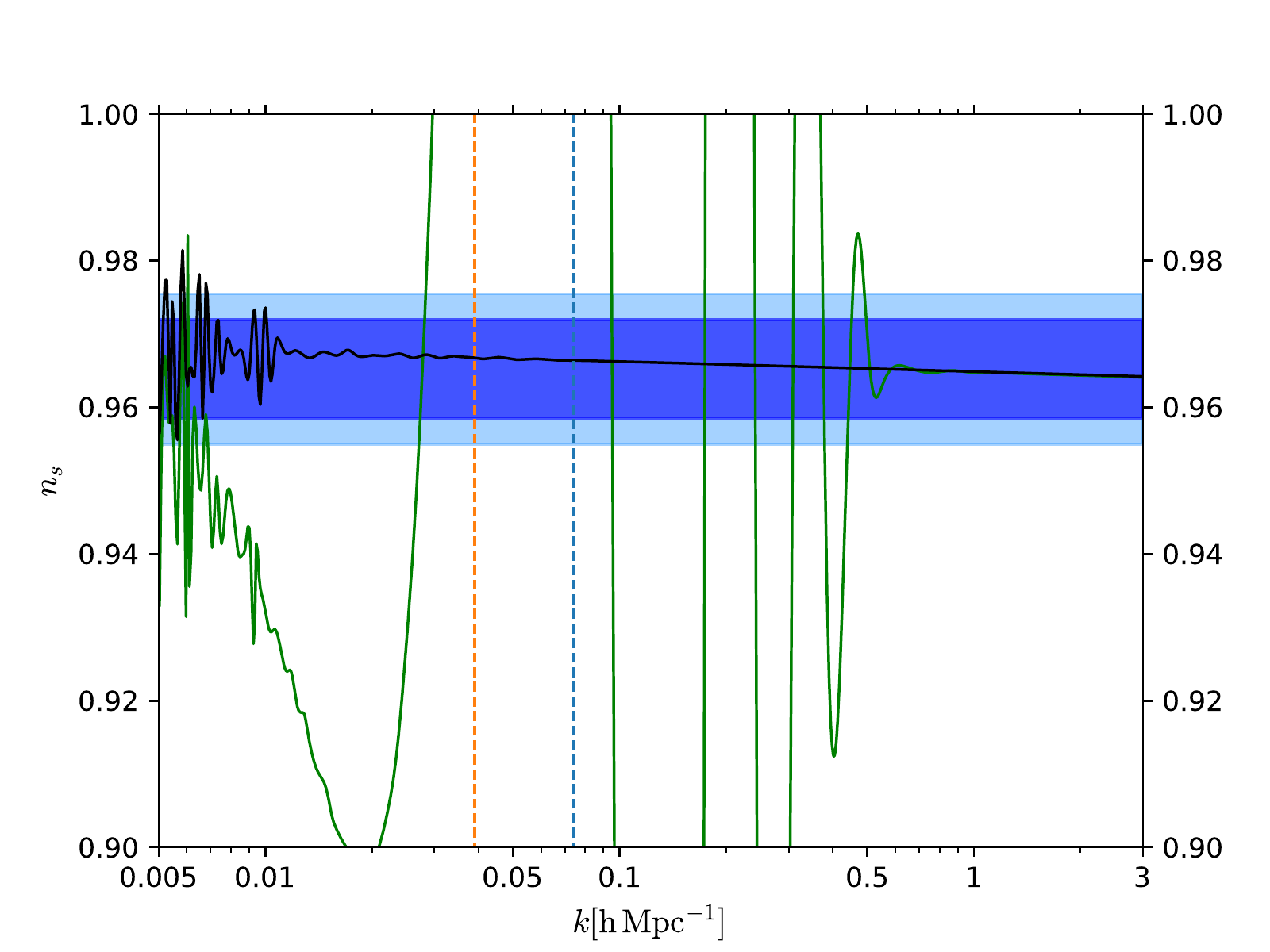}
\includegraphics[scale=0.548]{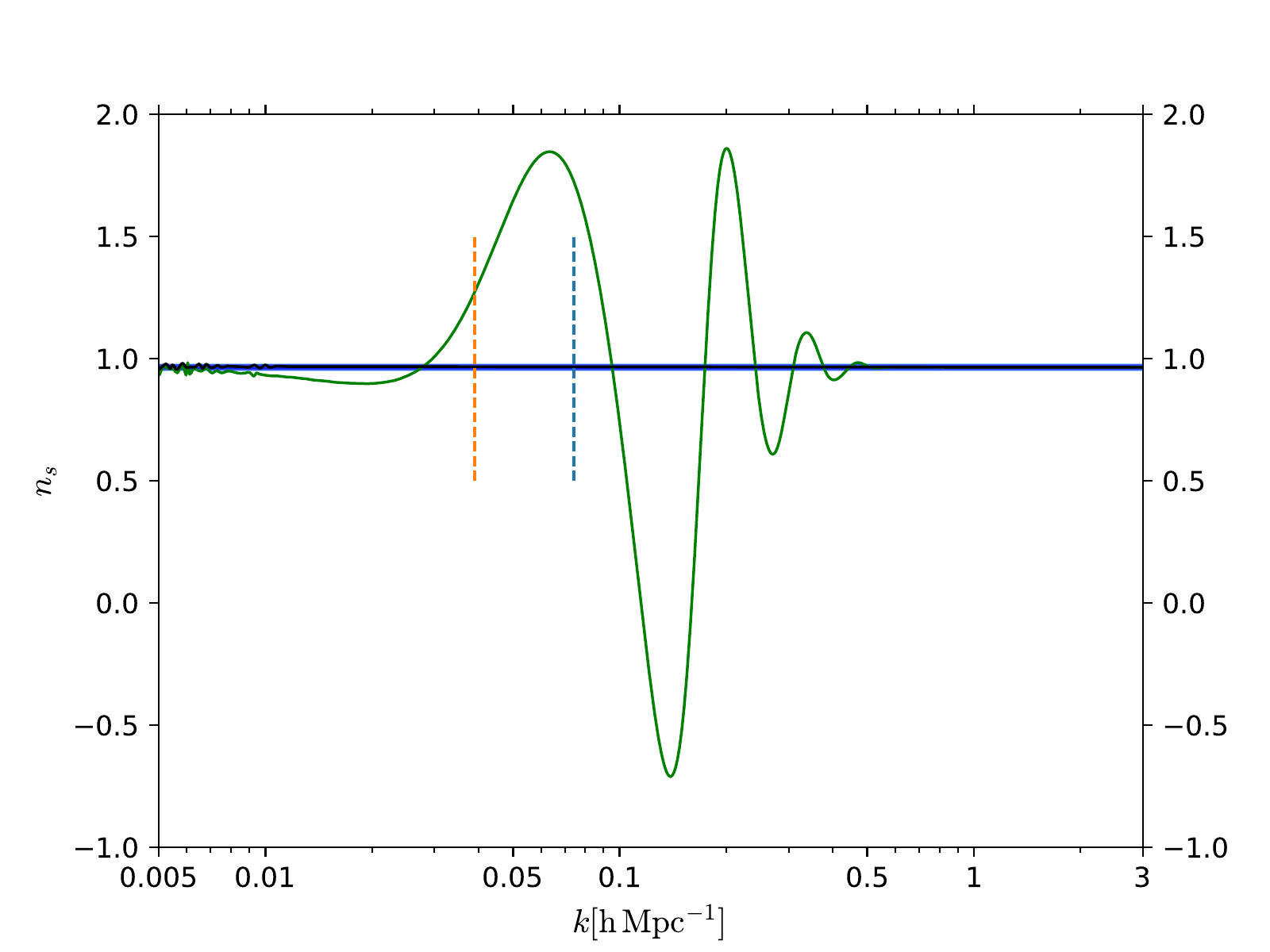} 
\caption{Evolution of the scalar spectral index $n_\mathrm{s}$ with respect to $k$, described by step (green) and quadratic chaotic (black) potentials for 71.0 e-folds of inflation. The orange vertical line corresponds to $k_{0}\simeq 0.04  \,h \,\rm Mpc^{-1}$, and the blue dashed line corresponds to the CMB observation by Planck Legacy pivot value $k_{\star}\simeq 0.0743 \,h \,\rm Mpc^{-1}$ ($k_{\star}\simeq 0.05 \rm \,Mpc^{-1}$)~\cite{akrami:2018b, Planck:2018jri}. The blue contours correspond to the $68\%$ and $95\%$ C.L. results from Planck 2018 TT,TE,EE+lowE+lensing data \cite{akrami:2018b, Planck:2018jri}.}\label{fig:CI_phys_units_ns_k}
\end{figure}

\begin{figure}[htbp] 
\includegraphics[scale=0.548]{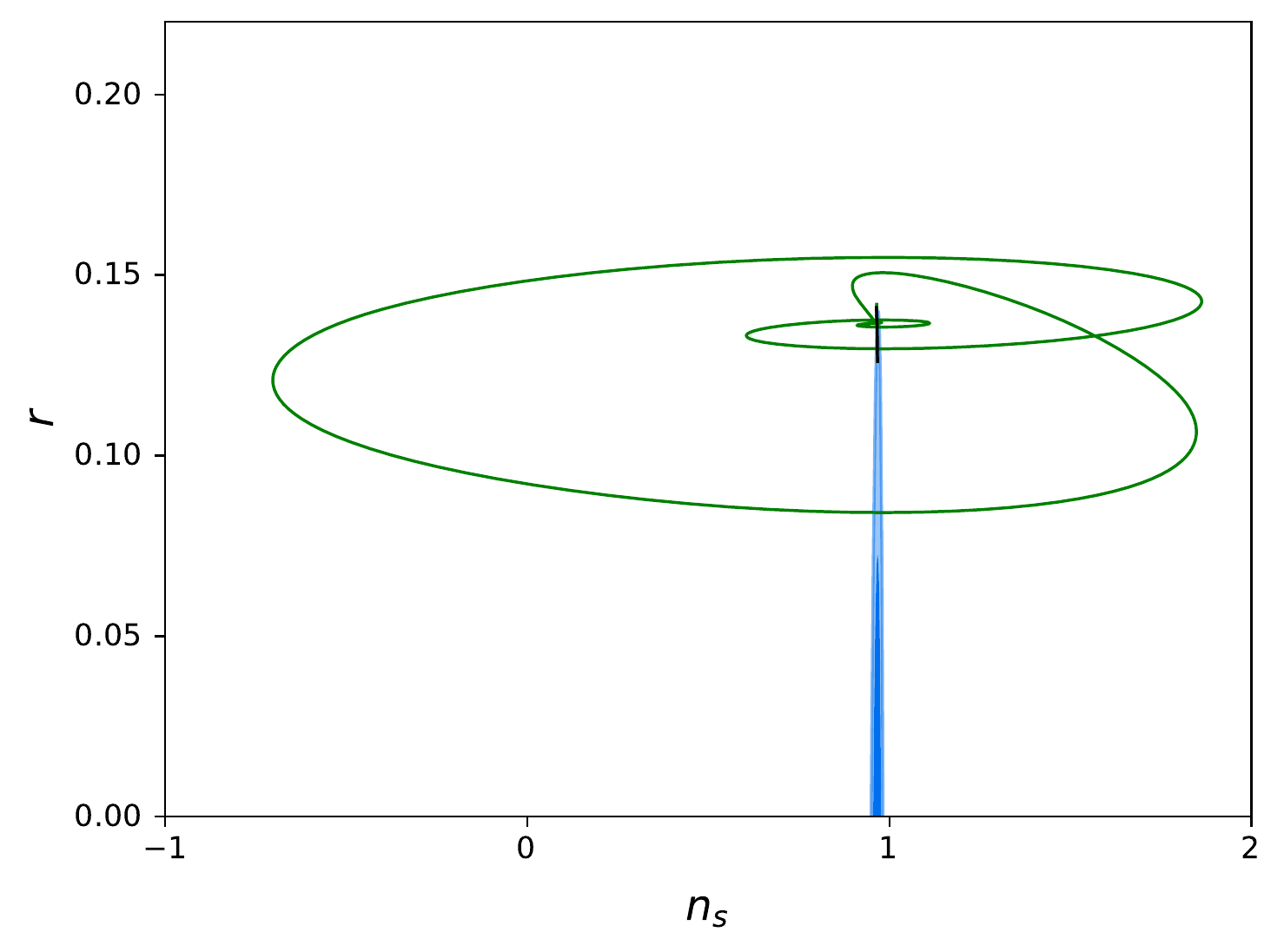}
\includegraphics[scale=0.548]{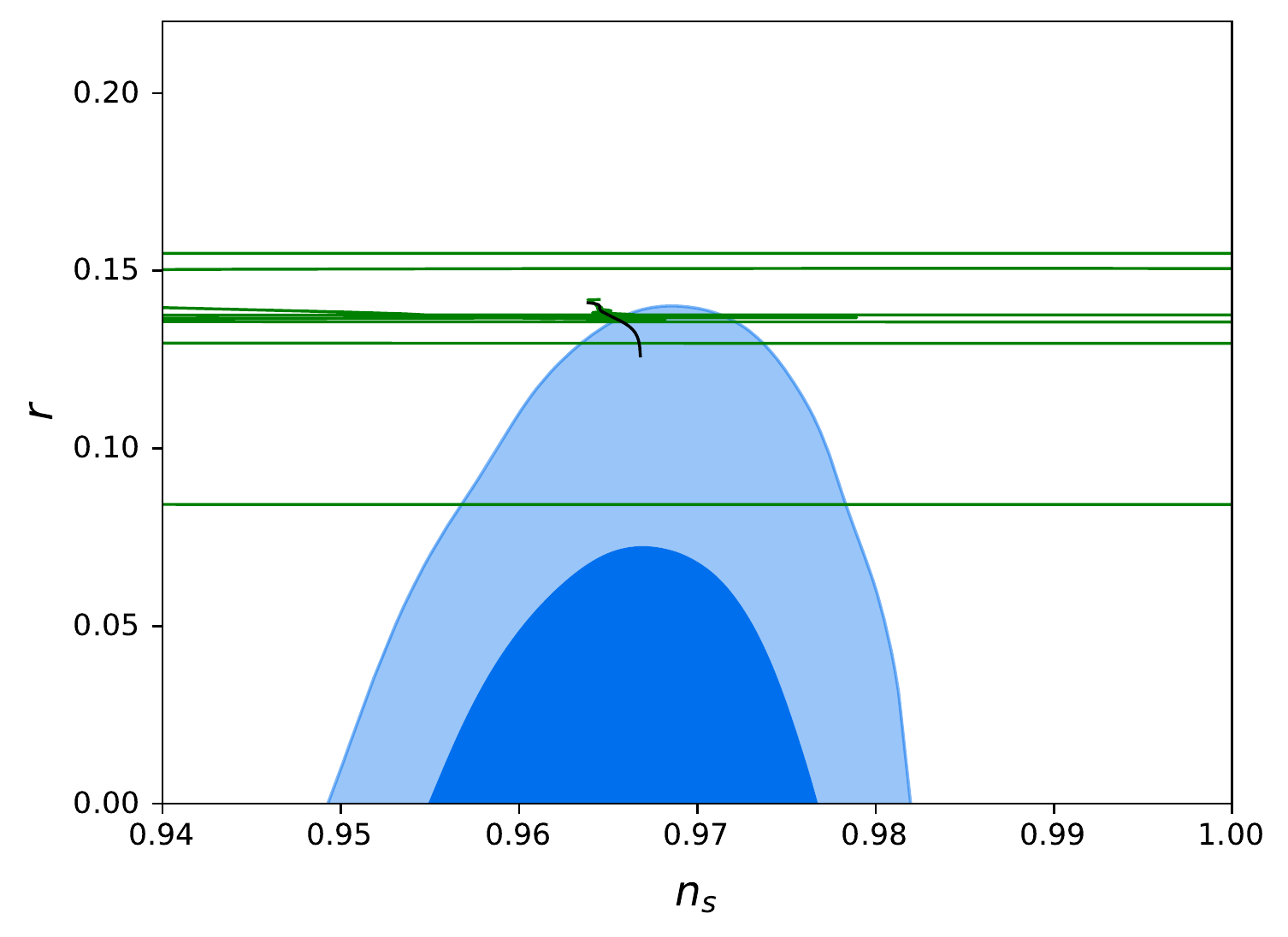} 
\caption{Observable parameters: scalar spectral index $n_\mathrm{s}$ and the tensor-to-scalar ratio $r$. This plot shows the allowed trajectories in the $(n_\mathrm{s},r)$ plane, described by step (green) and quadratic chaotic (black) potentials for 71.0 e-folds of inflation. The blue contours correspond to the $68\%$ and $95\%$ C.L. results from Planck 2018 TT,TE,EE+lowE+lensing data \cite{akrami:2018b, Planck:2018jri}.}\label{fig:CI_phys_units_ns_r}
\end{figure}

\begin{figure}[htbp] 
\includegraphics[scale=0.90]{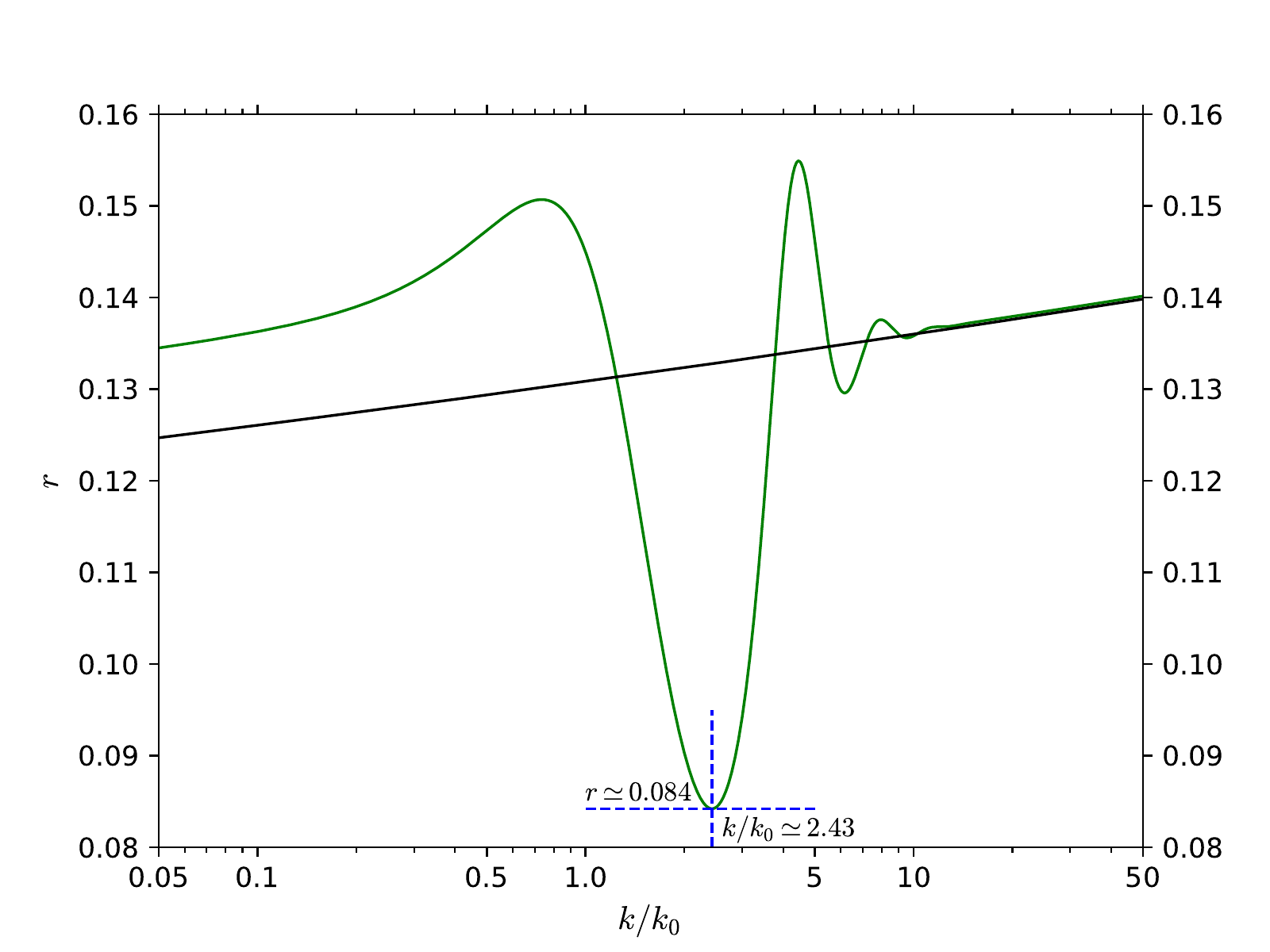}
\caption{Behaviour of the tensor-to-scalar ratio $r$ with respect to $k/k_{0}$, where $k_{0}\simeq 0.04 \,h \,\rm Mpc^{-1}$, described by step (green) and quadratic chaotic (black) potentials for 71.0 e-folds of inflation. Note that the minimum value $r\simeq 0.084$ corresponds to $k/k_{0}\simeq 2.43$. Indeed, this outcome lies within the Planck contours \cite{akrami:2018b}.}\label{fig:CI_r_k}
\end{figure}

In the following examples, we slightly change the initial parameters, since we set $P_{\mathcal{R}}\simeq 2.22\times 10^{-9}$ \cite{akrami:2018b} at $k_{0}\simeq 0.04 \,h \, \rm Mpc^{-1}$. Then, to normalise $P_{\mathcal{R}}$ with respect to the above value, we take $m=6.3\times 10^{-6}\, M_{Pl}$, $\phi_{\textnormal{step}}=15.5 M_{Pl}$, $c=0.0012$, $d=0.04  \, M_{Pl}$, and the initial value of the inflaton $\phi_{0}=16.8 \, M_{Pl}$. Now, inflation lasts $N_{e}=71.0$, but the step occurs at $\phi=\phi_{\textnormal{step}}$ around $N_{e}\simeq 10.5$. Therefore, once again, from the time of the step about 60 e-folds of inflation take place. The following plots figs.~\ref{fig:CI_PR_k}, \ref{fig:CI_phys_units_ns_k}, \ref{fig:CI_phys_units_ns_r}, and \ref{fig:CI_r_k} will show all relevant observable parameters. We perform a wide scan of the scale $k$ in terms of a pivot wave number $k_{0}$, and we numerically evaluate all quantities twice after the horizon crossing, since we check that indeed $P_{\mathcal{R}}$ and $P_{h}$ become constant at this point in time. 

Let us examine Fig.~\ref{fig:CI_PR_k}. First, the upshot of the power spectrum is in good agreement with the results of A. Gallego-Cadavid \cite{cadavid:2017}, where his conclusions indicate that a step in the potential can cause significant changes in the spectrum at scales around $k_{0}$. Additionally, we include the evolution of a featureless instance, in order to exemplify how the oscillatory proposal distances from the chaotic potential for approximately two decades $0.2< k/k_{0} < 20$; however, both $P_{\mathcal{R}}$'s behave the same before and after this range. Recall that we mentioned that perturbations around the step are not longer ``freeze in'' on super-horizon scales; this is true since $\mathcal{R}$ evolves after horizon crossing but only around the above regime ($0.2< k/k_{0} < 20$), therefore, this result implies that the oscillatory behaviour of $P_{\mathcal{R}}$, around the step, is due only to the evolution of the background, in particular because of the step potential. This feature background property was already pointed out in~\cite{Vallejo-Pena:2019lfo, Gordon:2000hv}. On the other hand, figs.~\ref{fig:CI_phys_units_ns_k}, \ref{fig:CI_phys_units_ns_r}, and \ref{fig:CI_r_k} bring new aspects of this model. We plot the evolution of the scalar spectral index $n_\mathrm{s}$ with respect to $k$; $n_\mathrm{s}$ versus the tensor-to-scalar ratio $r$; and $r$ versus $k$. As expected a quadratic chaotic scenario presents a nearly scale invariant $n_\mathrm{s}$ {\footnote{In order to compute the scalar spectral index $n_\mathrm{s}$, we implement a numerical code to compute the numerical derivative $n_\mathrm{s}-1\equiv \frac{d\ln P_{\mathcal{R}}}{d\ln k}$. We use Python~\cite{10.5555}: numpy.diff~\cite{harris2020array}; and scipy.interpolate~\cite{2020SciPy-NMeth}.}}, whilst the potential with a step yields an oscillatory $n_\mathrm{s}$; where it fluctuates from negative values to greater than one. This outcome certainly needs further analysis and is left for future work. In addition, more than one region of ($n_\mathrm{s},r$), see Fig.~\ref{fig:CI_phys_units_ns_r}, was brought back to the observation window. And in particular, from Fig.~\ref{fig:CI_r_k} one can compute the minimum of $r\simeq 0.084$ that corresponds to $k/k_{0}\simeq 2.43$; where this value indeed lies within the observable contours \cite{akrami:2018b, Planck:2018jri}.  


However, we ignored that to examine the hints of a dip and a bump in the spectrum of primordial perturbations, by means of the features of the potential, the range of scales $k = 0.002 \rm Mpc^{-1}$ and $k = 0.0035 \rm Mpc^{-1}$ are used~~\cite{DiValentino:2016ikp,Benetti:2016tvm,GallegoCadavid:2016wcz}. Due to the perceptive comments of the reviewer, we decide to examine a particular example with $k_0 = 0.0014 \,\rm Mpc^{-1}$ $(0.002 \,h \,\rm Mpc^{-1})$, and we scan the range $0.05 \leq k/k_0 \leq 50$ (see Appendix~\ref{appendix_a}). The general behaviours of $P_{\mathcal{R}}$ and $r$ are qualitatively the same as the results with $k_{0}= 0.0269 \rm Mpc^{-1}$ $(0.04 \,h \,\rm Mpc^{-1})$, and, consequently, $n_\mathrm{s}$ also presents the same outcome. However, as the reviewer remarked, the resulting observational values are, in fact, the consequence of the initial values of the parameters. So, for instance, the minimum value of $r$ is now $r\simeq 0.1$ at $k/k_{0}\simeq 2.48$ (see Fig.\ref{fig:appendix_CI_r_k_k0}). 

Notwithstanding, we analysed in great detail a CI model; in the next section, we will present an enhanced scenario: WI. 

\section{Warm Inflation}
%
To begin with, WI brings radiation into the evolution of inflation. Thus, a thermal bath imbibes the inflaton field, modifying the dynamics at both the background and the perturbation levels. Initially, the main input is to introduce a dissipative term $\Upsilon\dot{\phi}$ into the inflaton evolution equation as a source of radiation production \cite{Berera:1995wh,Berera:1995ie,Berera:1996fm,Berera:2008ar}. To describe this mechanism, all macroscopic motion must be slow relative to the relevant microscopic time scales, and then the macroscopic dynamics can be treated adiabatically. To ensure the process of near-equilibrium, as mentioned earlier, one must satisfy the adiabatic condition, namely $\Gamma/H>1$, where $\Gamma$ is the decay width, which quantifies microscopic interactions; and $H$ corresponds to a macroscopic measurement. Posteriorly, a noise force term was included to also drive inflaton fluctuations, with a fluctuation-dissipation theorem that uniquely specifies inflaton fluctuations \cite{Berera:1995wh,Berera:1995ie,Berera:1996fm,Berera:2008ar}. WI effectively explains the transition from the vacuum- to the radiation-dominated universe.; however, this framework yields some shortcomings. For instance, its realisation is due to non-equilibrium dissipative effects, and these are Boltzmann suppressed unless the particles in the radiation bath are relativistic, i.e., highly energetic. These relativistic particles typically induce large thermal corrections to the inflaton's mass that may prevent slow-roll unless the associated inflaton couplings are very suppressed; therefore, the thermal bath could become inefficient during inflation~\cite{Berera:1998gx,Yokoyama:1998ju}. Initially, the main solution to the problems mentioned above was to consider scenarios in which the inflaton only couples directly to heavy fields, which in turn decay into light particles in the thermal bath~\cite{Berera:2002sp}; however, even when thermal corrections to the inflaton potential remain under control, dissipative effects can only be significant if a large number of fields coupled to the inflaton are considered~\cite{Moss:2006gt,Bastero-Gil:2010dgy,Bastero-Gil:2012akf}. These large field multiplicities can be found in specific constructions, e.g. string theory~\cite{Bastero-Gil:2011zxb}; and in the same context the distributed-mass model (DM model)~\cite{Berera:1998gx, Berera:1999wt, Berera:1999ws, Berera:1998px} was proposed. Recently, a supersymmetric DM model \cite{Bastero-Gil:2018yen} has acquired important attention since, for instance, thermal corrections to the inflaton's potential are under control, and the number of supersymmetric fields can be small and sustain enough dissipation during inflation. In this work, we will implement the aforementioned DM scenario within the WI framework.  

We will start with the WI background evolution equations for the inflaton-radiation system, given by: 
\begin{eqnarray}
\ddot{\phi}+(3H+\Upsilon)\dot{\phi}+ V_{\phi} &=& 0 \,, \nonumber\\
\dot{\rho}_{r}+4H\rho_{r} &=& \Upsilon\dot{\phi}^{2} \,, \label{background_eqs}
\end{eqnarray}
where $\Upsilon=\Upsilon(T,\phi)$ is the dissipative coefficient{\footnote{Indeed, $\Upsilon$ can depend on either temperature $T$ or $\phi$ or both. For instance, a particular supersymmetric model yields a dissipative coefficient $\Upsilon \propto T^{3}/\phi^{2}$ \cite{Bastero-Gil:2009sdq}.}} in the leading adiabatic approximation, and it is computed from first principles provided the relevant interactions between the scalar field and the thermalised degrees of freedom (dof); $\rho_{r}=\pi^{2}g_{*}T^{4}/30$, $g_{*}$ being the effective no. of light dof. And once again dots correspond to time derivatives, $V_{\phi}$ is the derivative of the potential energy with respect to the field, and $H$ is the Hubble parameter, given by the Friedmann equation for a flat FRW universe:
\begin{equation}\label{Hubble}
3H^{2}=\frac{\rho}{M_{Pl}^{2}} \,,
\end{equation}
where $\rho=\rho_{\phi}+\rho_{r}$ is the total energy density, with $\rho_{\phi}=\dot{\phi}^{2}/2+V(\phi)$. In addition, the field pressure and the equation of state for the degrees of freedom of radiation are $p_{\phi}=\dot{\phi}^{2}/2-V(\phi)$ and $p_{r}=\rho_{r}/3$ respectively. The dissipation coefficient is given by \cite{Bastero-Gil:2018yen}:
\begin{equation}
\Upsilon = C_{\phi}\phi^{n}\,, \quad n\geq 1 \,,  
\end{equation}
where this expression comes from the dynamics of a supersymmetric DM model. This particular scheme arises from a general form of an effective $\rm N=1$ global SUSY theory version of the DM model with chiral superfields $\Phi$, $X_{i}$ and $Y_{i}$ \cite{Hall:2004zr, Bastero-Gil:2009sdq, Bastero-Gil:2010dgy, Bastero-Gil:2012akf}. The chiral superfields $\Phi$, $X_{i}$ and $Y_{i}$ have (scalar, fermion) components ($\phi$,$\psi_{\phi}$), ($\chi_{i}$,$\psi_{\chi_{i}}$) and ($\sigma_{i}$,$\psi_{\sigma_{i}}$), respectively. Thus, $g_{*}= 1 + 15 N_M/4$, $N_M$ being the no. of bosonic $\chi_i$ (fermionic $\psi_i$) light degrees of freedom at horizon crossing. There are more consistent with microscopic derivations of a temperature-independent $\Upsilon$ (see, e.g., refs.\cite{deOliveira:1997jt, delCampo:2007cy, delCampo:2010by, Ramos:2001zw, Ramos:2013nsa}). Furthermore, in the DM model, the average decay width $\Gamma \propto T$ \cite{Bastero-Gil:2018yen}. During inflation, the motion of the inflaton field must be overdamped to end the accelerated expansion, and this can be achieved due to either the Hubble rate, as in the CI case, or an extra friction term $\Upsilon$, or the interaction of both components. We can quantify this competition by defining the dissipative ratio $Q=\Upsilon/(3H)$. According to the ratio $Q$ we will have distinct cases: for $Q<1$, this is called weak dissipative warm inflation (WDWI); and when $Q\geq 1$ we are in strong dissipative warm inflation (SDWI). Importantly, given the presence of radiation during inflation, the potential acquires thermal corrections \cite{Bastero-Gil:2018yen}:
\begin{equation}\label{total-effective-potential-discrete}
V_{T}(T,\phi) = V_{T} \simeq  \frac{g^{2}T^{3}}{24}\phi \left[-\pi^{2}+3+\frac{4}{\pi}+\frac{3}{4\pi^{2}}(c_{b}-c_{f})\right] \,,
\end{equation}
where $g$ is a coupling constant, $c_{b}=5.41$ and $c_{f}=2.632$ are constants related to the contributions of bosson (b) and fermion (f) of the finite-temperature potential. This particular form of $V_{T}$ comes from a dissipative given by $\Upsilon\sim\phi$ \cite{Bastero-Gil:2018yen}. Note that $V_{T}$ could become relevant at background and fluctuation levels, as they adjust the slow-roll parameter $\epsilon_{H}$. However, in the scenario with a quartic chaotic potential $V(\phi)=\lambda\phi^4/4$ within the SDWI, such thermal corrections give a negligible contribution to both the effective potential and its first derivative in the perturbative regime \cite{Bastero-Gil:2018yen}. And we expect a similar outcome with a chaotic potential inflationary potential $V(\phi)=m^2\phi^2/2$, within the SDWI, since at the SDWI regime we typically find that $V(\phi)/M_{Pl}^{4} \sim 10^{-16}$ and $|V_{T}/M_{Pl}^{4}| \sim g^{2}\,10^{-23}$, therefore, for any value of the coupling $g$ in the perturbative regime we will have $|V_{T}/V(\phi)|\sim 10^{-7}$, and given that both $T$ and $\phi$ evolve nearly constant, until a few e-folds before inflation ends, the rate $|V_{T}/V(\phi)|$ will remain practically invariant during almost the entire period of inflation. Then thermal corrections to the inflaton potential do not considerably modify the full dynamics; hence, we will not consider these. Moreover, to describe a process of near-equilibrium, thermalisation must be satisfied, that is, the adiabatic condition $\Gamma/H>1$, which roughly translates into $T/H>1$ since $\Gamma \propto T$ \cite{Bastero-Gil:2018yen}. Without including the $T$-dependent corrections in the inflaton potential, we would have the standard relation $Ts_{r}=4\rho_{r}/3=4\pi^{2}g_{*}T^{4}/90$. Inflation lasts while the slow-roll parameter $\epsilon_{H}=-\dot{H}/H^{2}<1$, where $\dot{H} = -(\dot{\phi}^{2} + Ts_{r})/(2M_{Pl}^{2})$, so it ends when $\epsilon_{H}=1$. An inflationary scenario also requires a flat potential, and this condition is measured by the $\eta_{H}=-\ddot{\phi}/(H\dot{\phi})$ parameter. where a particular potential preserves the flatness having $|\eta_{H}|<1$.

Once again, we have to introduce the scalar perturbations; however, for WI, these are also accompanied by thermal fluctuations. Hence, during WI we have a multicomponent fluid: a mixture of a scalar field (the inflaton) interacting with the radiation fluid. The equations for the perturbations in a multicomponent fluid can be found, for example, in \cite{Kodama:1984ziu, Hwang:1991aj, Malik:2002jb}. Working in momentum space, defining the Fourier transform with respect to the comoving coordinates, the equations of motion for the radiation and momentum fluctuations with the comoving wavenumber $k$ are given by \cite{Hwang:1991aj}:
\begin{eqnarray} 
&& \dot{\delta\rho_{r}} + 4H\delta\rho_{r} = 4\rho_{r}\dot{\alpha}  + \frac{k^{2}}{a^{2}}\Psi_{r} + \delta Q_{r} + Q_{r}\alpha\,, \label{delta_rho_r}\\
&& \dot{\Psi}_{r} + 3H\Psi_{r} = -\frac{\delta\rho_{r}}{3} -\frac{4\rho_{r}}{3}\alpha - \Upsilon\dot{\phi}\delta\phi \,, \label{Psi_r}
\end{eqnarray}   
where $Q_{r}=\Upsilon\dot{\phi}^{2}$, and $\delta Q_{r}= \delta\Upsilon\dot{\phi}^{2}+2\Upsilon\dot{\phi}\dot{\delta\phi}-2\alpha\Upsilon\dot{\phi}^{2}$. We have also used the equation of state for the degrees of freedom of radiation $\delta p_{r} = \delta\rho_{r}/3$. For a scalar field that interacts with a fluid, the evolution equation for field fluctuations $\delta\phi$, which is described by a stochastic evolution determined by the Langevin-like equation \cite{Berera:1995wh, Berera:1999ws, Hall:2003zp, Calzetta:1986cq}:
\begin{equation}\label{deltaphi_eq}
\ddot{\delta\phi} + (3H+\Upsilon)\dot{\delta\phi} + \left(\frac{k^{2}}{a^{2}}+V_{\phi\phi}  \right)\delta\phi = (2(\Upsilon+H)T)^{1/2}a^{-3/2}\Xi_{k} -\delta\Upsilon\dot{\phi} + 4\dot{\alpha}\dot{\phi} + (2\ddot{\phi}+6H\dot{\phi}+\Upsilon\dot{\phi})\alpha  \,,
\end{equation}
where $\Xi_{k}\equiv \Xi(\mathbf{k},t)$ is a stochastic source that can be well approximated by a localised Gaussian distribution with correlation function given by:
\begin{equation}
\langle \Xi(\mathbf{k},t) \Xi(\mathbf{k}',t')\rangle = \delta(t-t')\delta^{(3)}(\mathbf{k}-\mathbf{k}') \,.
\end{equation}
And the evolution equation for the metric variable $\alpha$ is given by \cite{Hwang:1991aj}:
\begin{equation}\label{alpha_WI}
\ddot{\alpha} + 4H\dot{\alpha} + (3H^{2}+2\dot{H})\alpha = \frac{1}{2M_{Pl}^{2}}\left(\dot{\phi}\dot{\delta\phi}-\dot{\phi}^{2}\alpha-V_{\phi}\delta\phi+\frac{\delta\rho_{r}}{3}\right)\, .
\end{equation}
Given that a particular gauge choice would depend on the problem at hand, hence, in order to avoid any subsequent miscalculation, at linear order, following \cite{Hwang:1991aj} we introduce the gauge-invariant field, energy density, and momentum perturbations:
\begin{eqnarray}
&& \delta\phi^{GI} = \delta\phi + \frac{\dot{\phi}}{H}\alpha \,, \qquad \Psi_{\phi}^{GI} = \Psi_{\phi} - \frac{\rho_{\phi} +p_{\phi}}{H}\alpha \,. \\
&&\delta\rho_{r}^{GI} = \delta\rho_{r} + \frac{\dot{\rho}_{r}}{H}\alpha \,, \qquad  \Psi_{r}^{GI} = \Psi_{r} - \frac{\rho_{r}+p_{r}}{H}\alpha \,,
\end{eqnarray}
with $\Psi_{\phi}=-\dot{\phi}\delta\phi$. For a multicomponent fluid, the total gauge-invariant comoving curvature perturbation $\mathcal{R}$ is given by the following:
\begin{equation}
\mathcal{R} = -\frac{H}{\rho + p}\Psi_{T}^{GI} = -\frac{H}{\rho + p}\sum_{x}\Psi_{x}^{GI} = \sum_{x}\frac{\rho_{x} + p_{x}}{\rho + p}\mathcal{R}_{x} \,,
\end{equation}
where $\Psi_{T}^{GI}=\Psi_{\phi}^{GI}+\Psi_{r}^{GI}$, then $\rho= \rho_{\phi}+\rho_{r}$ is the total energy density, while $p = p_{\phi} + p_{r}$ is the total pressure, therefore $\rho + p = \dot{\phi}^{2} + 4\rho_{r}/3$. Each fluid is labelled with a subindex ``$x$'':
\begin{equation}
\mathcal{R}_{x}  = -\frac{H}{\rho_{x} + p_{x}}\Psi_{x}^{GI} \,.
\end{equation}
For instance, with a scalar field, we have $\Psi_{\phi}^{GI}=-\dot{\phi}\delta_{\phi}^{GI}$, which yields:
\begin{equation}
\mathcal{R}_{\phi}  = \frac{H}{\dot{\phi}}\delta\phi^{GI} = \alpha + \frac{H}{\dot{\phi}}\delta\phi \,.
\end{equation}
Since we work in a gauge-invariant domain, the primordial curvature perturbation has the property to be constant within few Hubble times after the horizon crossing ($k_{*}= a_{*}H_{*}$), therefore we can compute it at horizon exit and remain ignorant about the subhorizon physics during and after reheating until horizon re-entry of a given $\mathcal{R}$-mode{\footnote{$\mathcal{R}$ becomes constant on super-horizon scales; perturbations with comoving wave number $k$ are said to ``freeze in'' as soon as the comoving Hubble horizon shrinks so far that $k^{-1}>(aH)^{-1}$.}}.  
The power spectrum of the comoving curvature perturbation $\mathcal{R}$ is defined as:
\begin{equation}
P_{\mathcal{R}} = \frac{k^{3}}{2\pi^{2}}\left|\mathcal{R}\right|^{2} = \frac{k^{3}}{2\pi^{2}}\left(\frac{H}{p+\rho}\right)^{2}\left|\Psi_{T}^{GI}\right|^{2} \,.
\end{equation}
As an example of a scalar field, the power spectrum of the comoving curvature perturbation $\mathcal{R}_{\phi} $ is given by:
\begin{equation}
P_{\mathcal{R}_{\phi}} = \frac{k^{3}}{2\pi^{2}}\left(\frac{H}{\dot{\phi}}\right)^{2}\left|\delta\phi^{GI}\right|^{2} \,.
\end{equation}
Then from the amplitude of the curvature power spectrum we may determine the scalar spectral index $n_\mathrm{s}$ and the tensor-to-scalar ratio $r$. Moreover, since $T\ll M_{Pl}$ gravitational waves are not significantly affected by thermal effects, therefore, the primordial tensor spectrum $P_{h}$ is given in the standard inflationary form. Numerically, we evaluate $n_\mathrm{s}$ and $r$ when $P_{\mathcal{R}}$ and $P_{h}$ become constant, approximately three times after horizon crossing. To compute $n_\mathrm{s}$ once again, we use Python~\cite{10.5555}: numpy.diff~\cite{harris2020array}; and scipy.interpolate~\cite{2020SciPy-NMeth}. In the next subsection, we present an example of background and linear perturbation dynamics within a WI scenario.  

\subsection{WI results:}
To study the effects of features on the spectrum of primordial perturbations, we utilise a linear dissipative coefficient: 
\begin{equation}\label{linear_dc}
\Upsilon = C_{\phi}\phi \,. 
\end{equation}
We consider the above component since there are certain favourable features already published. First, this example works correctly without adding thermal contributions to the potential, provided that the result comes from a SDWI \cite{Bastero-Gil:2018yen}, which is this case of study. Second, the adiabatic condition $\Gamma/H>1$ or $T/H>1$ is easily achieved in the SDWI; so there is no need to examine this thoroughly. Last but not least, two swampland criteria, relevant for inflationary theories, have been intensively discussed \cite{Obied:2018sgi,Agrawal:2018own}, $|\Delta\phi|/M_{Pl}<\Delta$ and $M_{Pl}|V_{\phi}|/V>c$, provided that $V>0$, where $\{\Delta,c\}\sim\mathcal{O}(1)$. And these criteria are very consistent with a WI paradigm described by a DM model \cite{Bastero-Gil:2018yen}. In fact, the dissipative characteristic in WI was observed to inherently make it compatible with the swampland criteria, as already noted in the literature \cite{Das:2018hqy,Das:2018rpg,Motaharfar:2018zyb,Yi:2018dhl,Lin:2018edm}. Numerically, we take $m=3.09897\times 10^{-8} M_{Pl}$, $\phi_{\textnormal{step}}=0.778 \,M_{Pl}$, $c=0.012$, $d=0.04 \, M_{Pl}$, $C_{\phi}=1.5\times 10^{-5}$, $N_{M}=10$, therefore $g_{*}=77/2$. In addition, the initial value of the inflaton $\phi_{0}=0.845 \, M_{Pl}$. Inflation lasts $N_{e}=70.2$. The step occurs at $\phi=\phi_\textnormal{step}$ around $N_{e}\simeq 10.1$. However, note that the parameter $c$ is one order of magnitude larger than one in the CI case. We ran a few examples with $c=0.0012$, yet the potential feature was not noticeable. Therefore, we decided to increase $c$ to show such distinctive characteristics of $V(\phi)$. In addition, given the aforementioned set of input values, we guarantee that the system is within a SDWI regime; where, in fact, the parameter $Q_{*}\geq 390$.  

\begin{figure}[htbp] 
\includegraphics[scale=0.548]{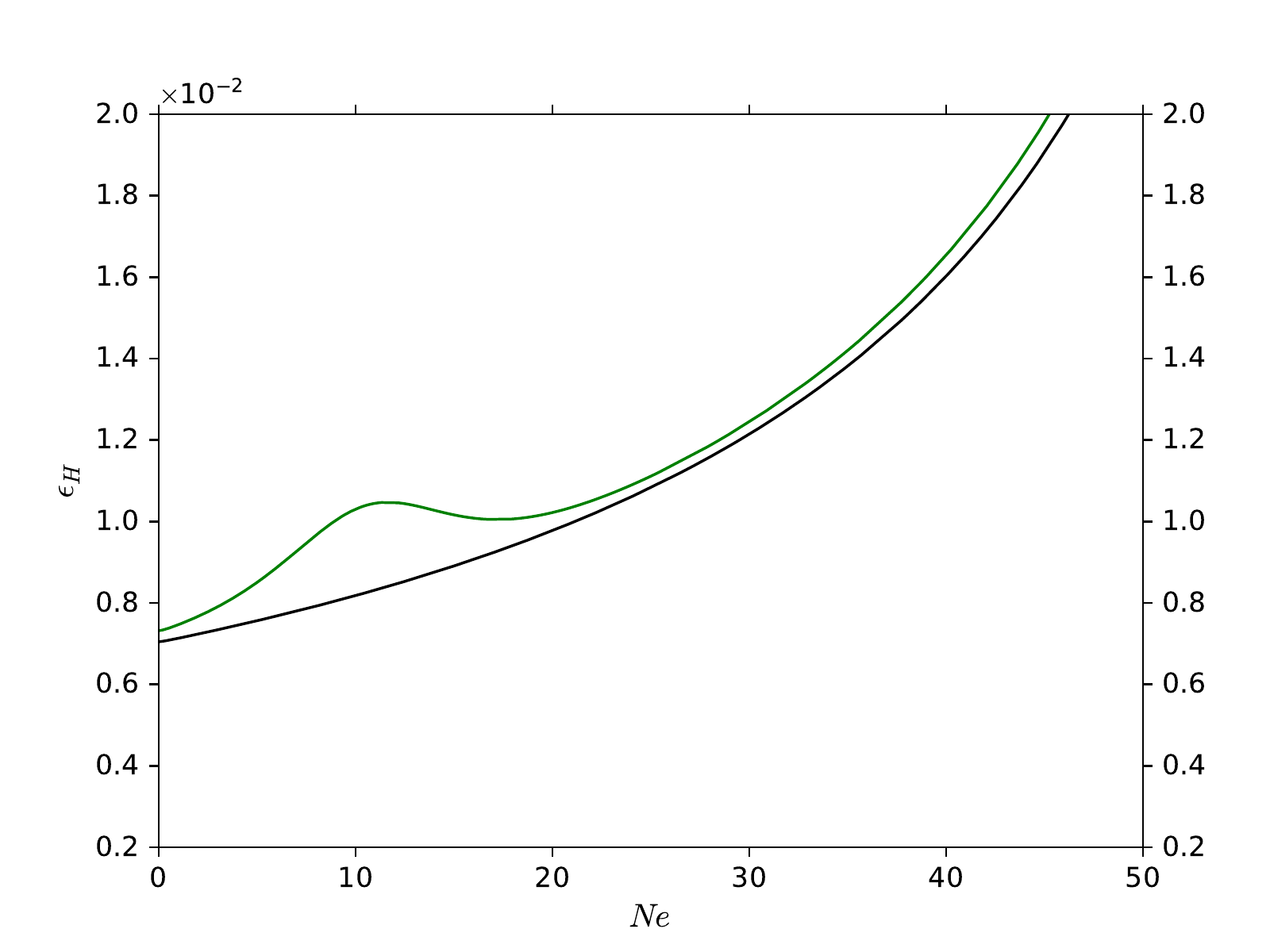}
\includegraphics[scale=0.548]{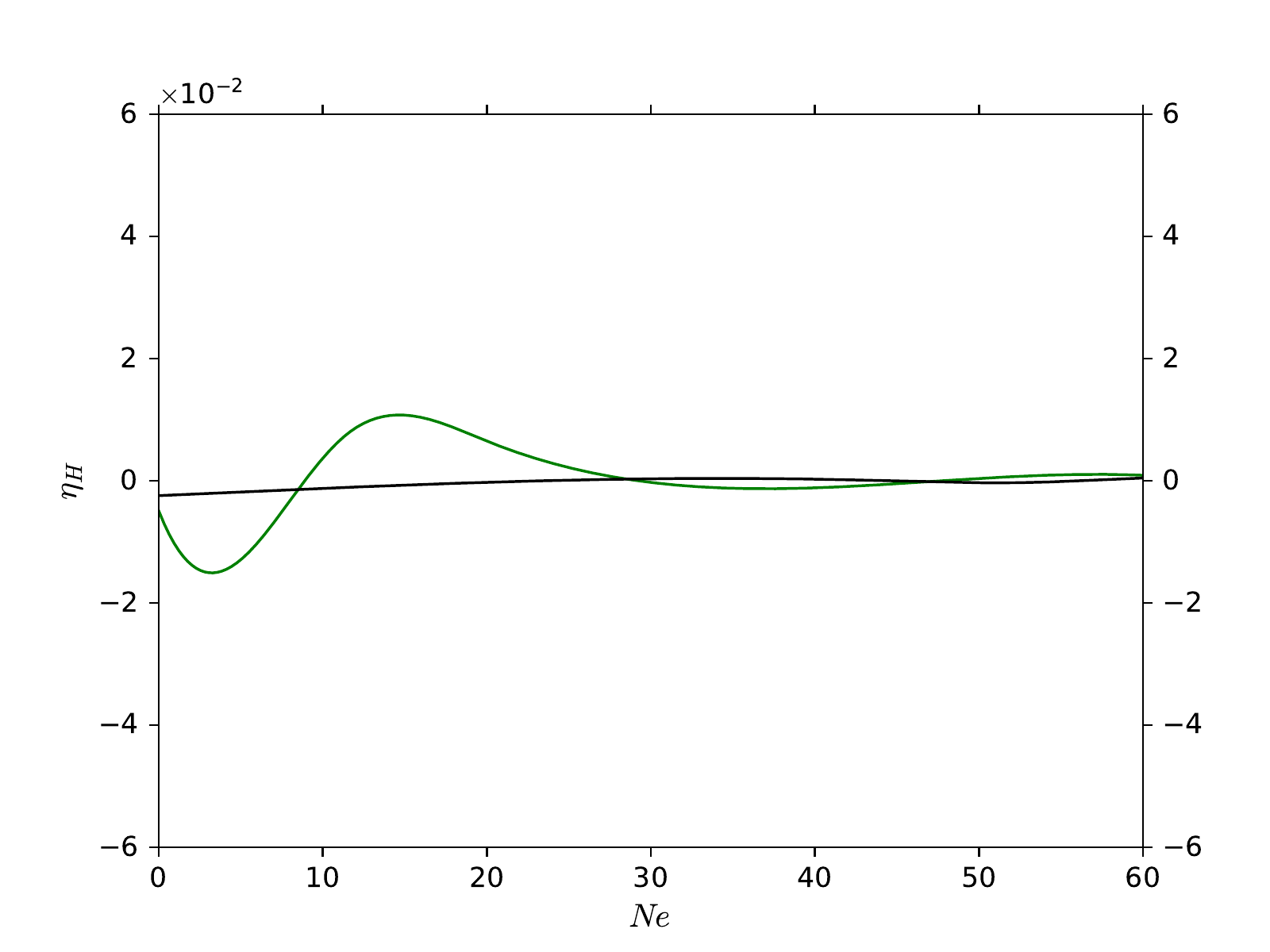} 
\caption{Behaviour of the slow-roll parameters $\epsilon_{H}$ and $\eta_{H}$ with respect to the number of e-folds $N_{e}$ within the WI scheme, described by step (green) and quadratic chaotic (black) potentials for 70.2 e-folds of inflation. Note that the step occurs at $N_{e}\simeq 10$.}\label{fig:WI_slow_roll_parameters}
\end{figure}

In Fig.~\ref{fig:WI_slow_roll_parameters} we show the effects of the features on the slow-roll parameters with respect to the number of e-foldings $N_{e}$. The black line characterises a quadratic chaotic description, whilst the green line corresponds to a step potential. Once again, the feature occurs around $N_{e}\simeq 10$. Contrary to the CI case, the contrast between both potentials becomes more evident for longer number of e-folds; in fact, note how the $\eta_{H}$ parameter (right hand of Fig.~\ref{fig:WI_slow_roll_parameters}) oscillates more than the CI example. Actually, this outcome will be determinant when analysing the linear perturbations, since we will numerically evaluate $P_{\mathcal{R}}$ and $P_{h}$ approximately three times after horizon crossing, where they certainly become constant.

\begin{figure}[htbp] 
\includegraphics[scale=0.548]{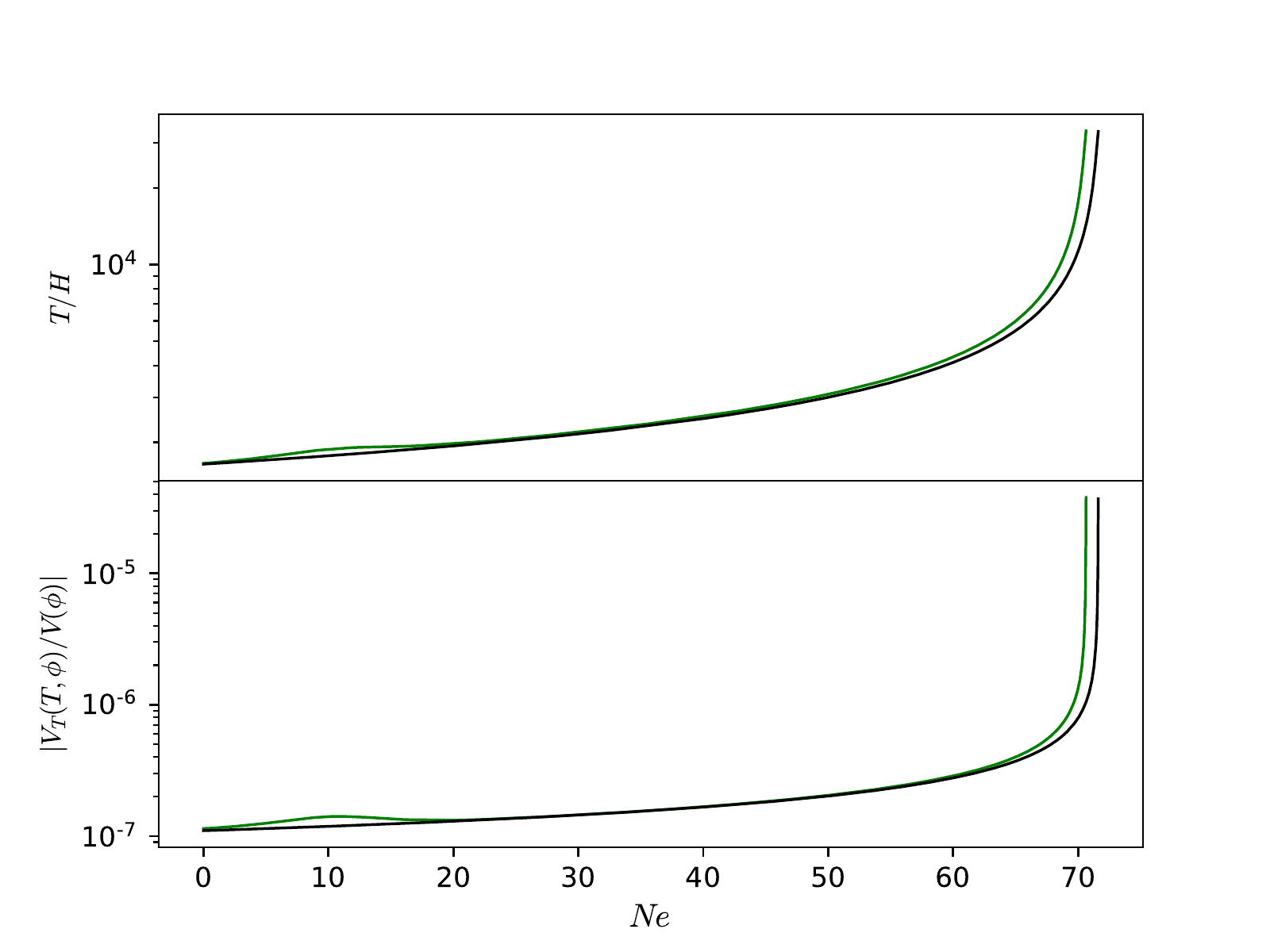}
\includegraphics[scale=0.548]{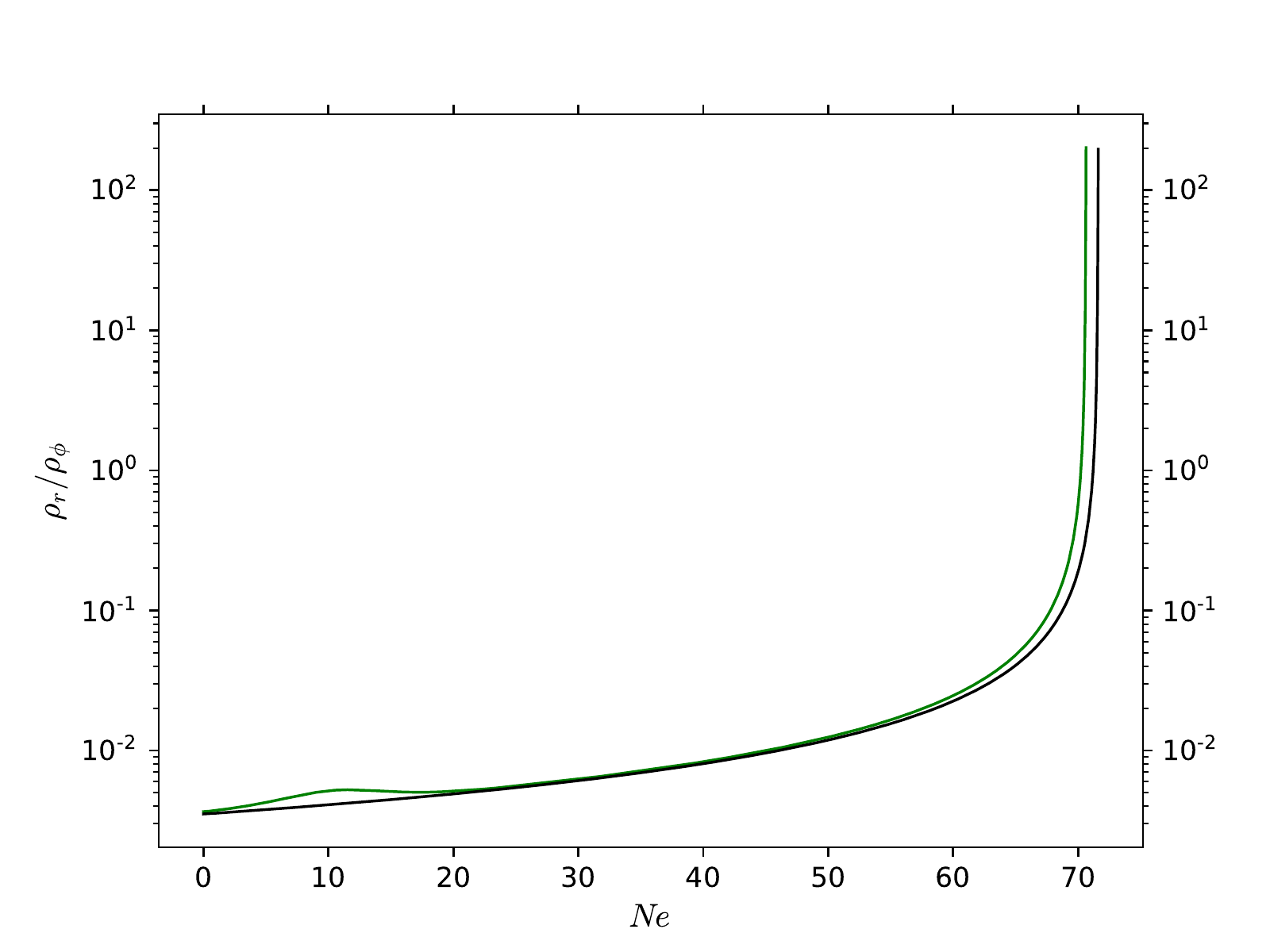} 
\caption{Left plot: evolution of the rates $T/H$ and $|V_{T}/V(\phi)|$ (for the coupling constant $g=1$) with respect to the number of e-folds $N_{e}$. Left plot: evolution of $\rho_{r}/\rho_{\phi}$ with respect to the number of e-folds $N_{e}$. In both panels, WI is described by step (green) and quadratic chaotic (black) potentials for 70.2 e-folds of inflation. Note that the step occurs at $N_{e}\simeq 10$. Remarkably $T/H\gg 1$ and $10^{-7}<|V_{T}/V(\phi)|<10^{-4}$, so thermal corrections to the potential are under control. Moreover, as expected, $\rho_{R}/\rho_{\phi}$ increases since the inflaton energy dissipates into radiation; however, for the step potential at $N_{e}\simeq 10$ it presents a sudden bump that slowly dies away.}\label{fig:WI_conditions}
\end{figure}

Then in Fig.~\ref{fig:WI_conditions} we show the behaviour of the rates $T/H$ and $|V_{T}/V(\phi)|$. Note that the adiabatic condition, $T/H>1$, is easily satisfied given that $T/H \sim 10^{3}$. On the other hand, thermal corrections to the inflaton's potential are not relevant in the SDWI regime, that is, $10^{-7}<|V_{T}/V(\phi)|<10^{-4}$. Furthermore, as the inflaton evolves, energy dissipates into radiation, which can be seen on the right-hand side of Fig.~\ref{fig:WI_conditions}; however, for the featured potential (green line), around the step $N_{e}\simeq 10$, a sudden bump occurs, but this slowly disappears, and from there both potentials yield similar results.  

\begin{figure}[htbp] 
\includegraphics[scale=0.90]{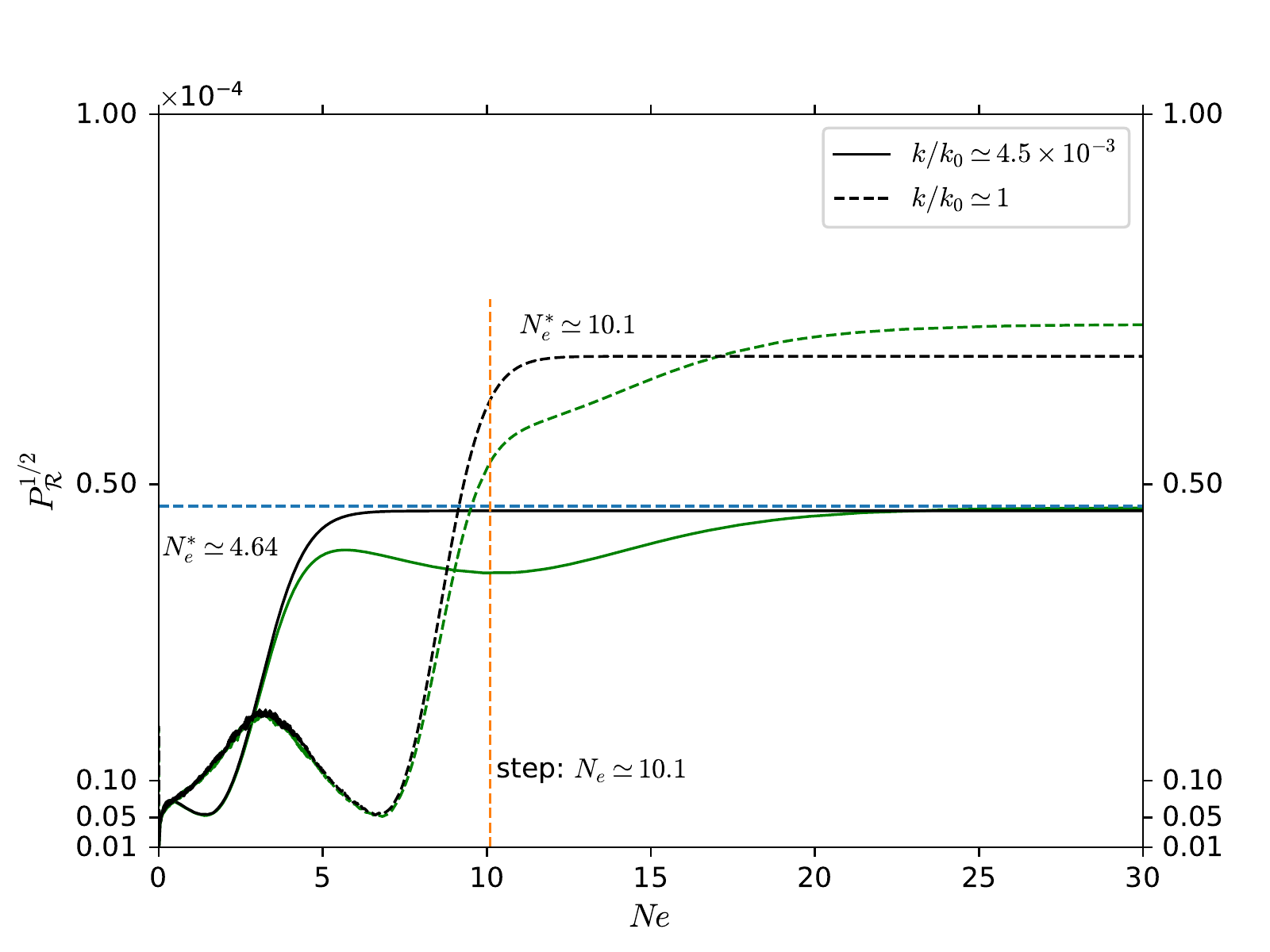}
\caption{Behaviour of the square root of the curvature power spectra $P_{\mathcal{R}}^{1/2}$ with respect to the number of e-folds $N_{e}$, described by step (green) and quadratic chaotic (black) potentials for $70.2$ e-folds of inflation. We show the outcome for two different wave numbers $k/k_{0} \simeq 1, 4.5\times 10^{-3}$, where $k_{0}\simeq 1.3\times 10^{-3}\,h\,\rm Mpc^{-1}$. Solid lines correspond to $N_{e}^{*}\simeq 4.64$, whilst dashed lines correspond to $N_{e}^{*}\simeq 10.1$. The blue dashed line corresponds to the CMB observations by the Planck Legacy value $P_{\mathcal{R}}^{1/2}\simeq 4.7\times 10^{-5}$ \cite{akrami:2018b}.}\label{fig:WI_PR}
\end{figure}

Then, Fig.~\ref{fig:WI_PR} shows the behaviour of the square root of the curvature power spectra $P_{\mathcal{R}}^{1/2}$ with respect to $N_{e}$. We only present the outcome of two distinct scales $k/k_{0} \simeq 1, 4.5\times 10^{-3}$, where $k_{0}\simeq 1.3\times 10^{-3}\,h\,\rm Mpc^{-1}$. First, a straightforward result is the growth of the amplitude of the power spectrum before horizon crossing, this is due to the stochastic source $\Xi_{k}\equiv \Xi(\mathbf{k},t)$. Then, this figure helps us to illustrate how the step spectrum behaves compared with the featureless one. Note that in the solid-lined example $P_{\mathcal{R}}^{1/2}$ changes smoothly when the step occurs, in contrast to CI. On the other hand, in the second case $k/k_{0} \simeq 1$ (dashed line), the step enhances the evolution of $P_{\mathcal{R}}^{1/2}$; however, in the following plots, we show that both potentials, in fact, produce quite similar results, which could be an indication of the influence of WI. Once again a discrepancy arises between the horizon crossing value and that of $P_{\mathcal{R}}^{1/2}$ constant comes up. This shift occurs due to a combined effect from the background (as in CI)~\cite{Vallejo-Pena:2019lfo, Gordon:2000hv} and entropy perturbations~\cite{DeOliveira:2001he,Gordon:2000hv,Wands:2000dp}, and both take place around the feature scale $k_{0}$. 


%

Recall that we have numerically integrated the background and perturbation equations. The set of chosen parameters guarantees that WI occurs in the SDWI regime; moreover, we scan the scale $k$ in terms of a pivot wave number $k_{0}\simeq 1.3\times 10^{-3}\,h\,\rm Mpc^{-1}$, and we numerically evaluate all quantities three times after horizon crossing, since we check that $P_{\mathcal{R}}$ and $P_{h}$ are indeed constant at this point in time. In addition, in the stochastic Langevin evolution equation for the inflaton field perturbation, eq.~(\ref{deltaphi_eq}), the stochastic noise term is numerically implemented in the time discretised code with an amplitude given by:
\begin{equation}
\Xi_{k} = \frac{\cal{G}}{\sqrt{dt}}
\end{equation}
where $\cal{G}$ are random numbers obtained from a zero-mean unit-variance Gaussian distribution~\cite{10.5555/1403886}. Due to the noise average of the power spectrum having to be taken over a considerable number of random realisations, this implementation requires very expensive computational time. Systematically, we executed two different quantities of numerical runs: 500 random numbers for $0.01\leq k/k_{0}\leq 32.5$; and 50 random numbers for $35\leq k/k_{0}<100$. 

%
%

\begin{figure}[htbp] 
\includegraphics[scale=0.90]{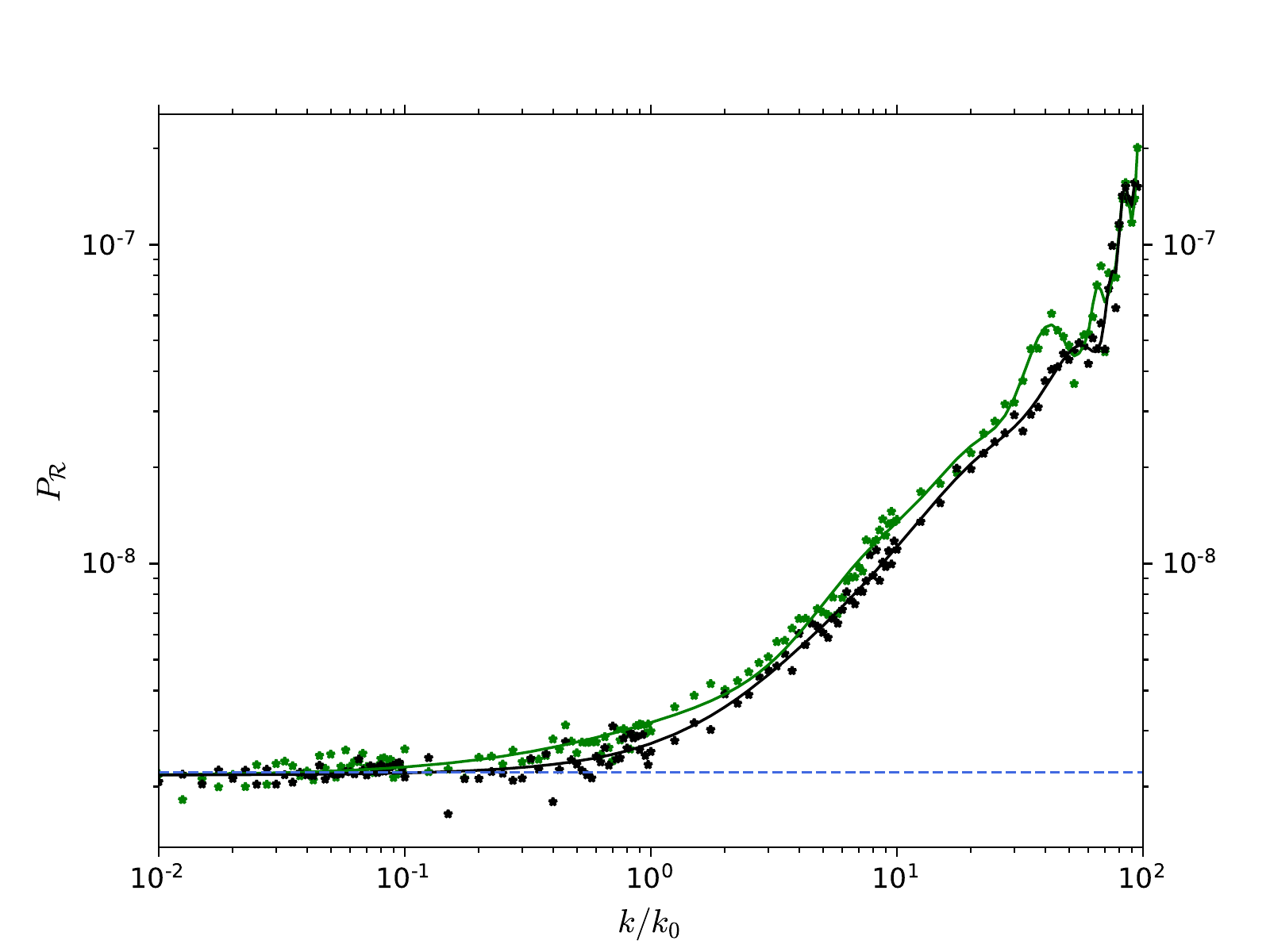}
\caption{Behaviour of the curvature power spectra $P_{\mathcal{R}}$ with respect to the ratio $k/k_{0}$, where $k_{0}\simeq 1.3\times 10^{-3}\,h\,\rm Mpc^{-1}$, described by step (green) and quadratic chaotic (black) potentials for $70.2$ e-folds of inflation. Filled stars are numerical values; while solid lines are the result of interpolation using a Python algorithm for curve fitting with spline functions: scipy.interpolate.splrep~\cite{2020SciPy-NMeth}. The blue dashed line corresponds to the CMB observations by the Planck legacy value $P_{\mathcal{R}}\simeq 2.22\times 10^{-9}$ \cite{akrami:2018b}.}\label{fig:WI_PR_k}
\end{figure}

Our numerical results for $P_{\mathcal{R}}=P_{\mathcal{R}}(k/k_{0})$ are shown in Fig.~\ref{fig:WI_PR_k}. Filled stars are our numerical results; while solid lines are the result of interpolation using a Python algorithm for curve fitting with spline functions: scipy.interpolate.splrep~\cite{2020SciPy-NMeth}. This method works by introducing a set of data points and then determining a smooth spline approximation of degree $\cal{K}$. The relevant parameters are: a smoothing condition $\cal{S}$; and the degree of spline fit $\cal{K}$. In our results, we use: ${\cal{K}}=5$ and ${\cal{S}}=1\times 10^{-15}$. Clearly $P_{\mathcal{R}}$ rises up for smaller scales, and this growth is, in fact, present in both potentials. The perturbations are amplified by WI. Moreover, small oscillations occur much later after the scale $k/k_{0}=1$, that is, at $k/k_{0}\sim 40$; however, our results do not entirely guarantee that these variations occur only on the featured potential. Further investigation beyond $k/k_{0}> 100$ can shed light on this matter.   

\begin{figure}[htbp] 
\includegraphics[scale=0.548]{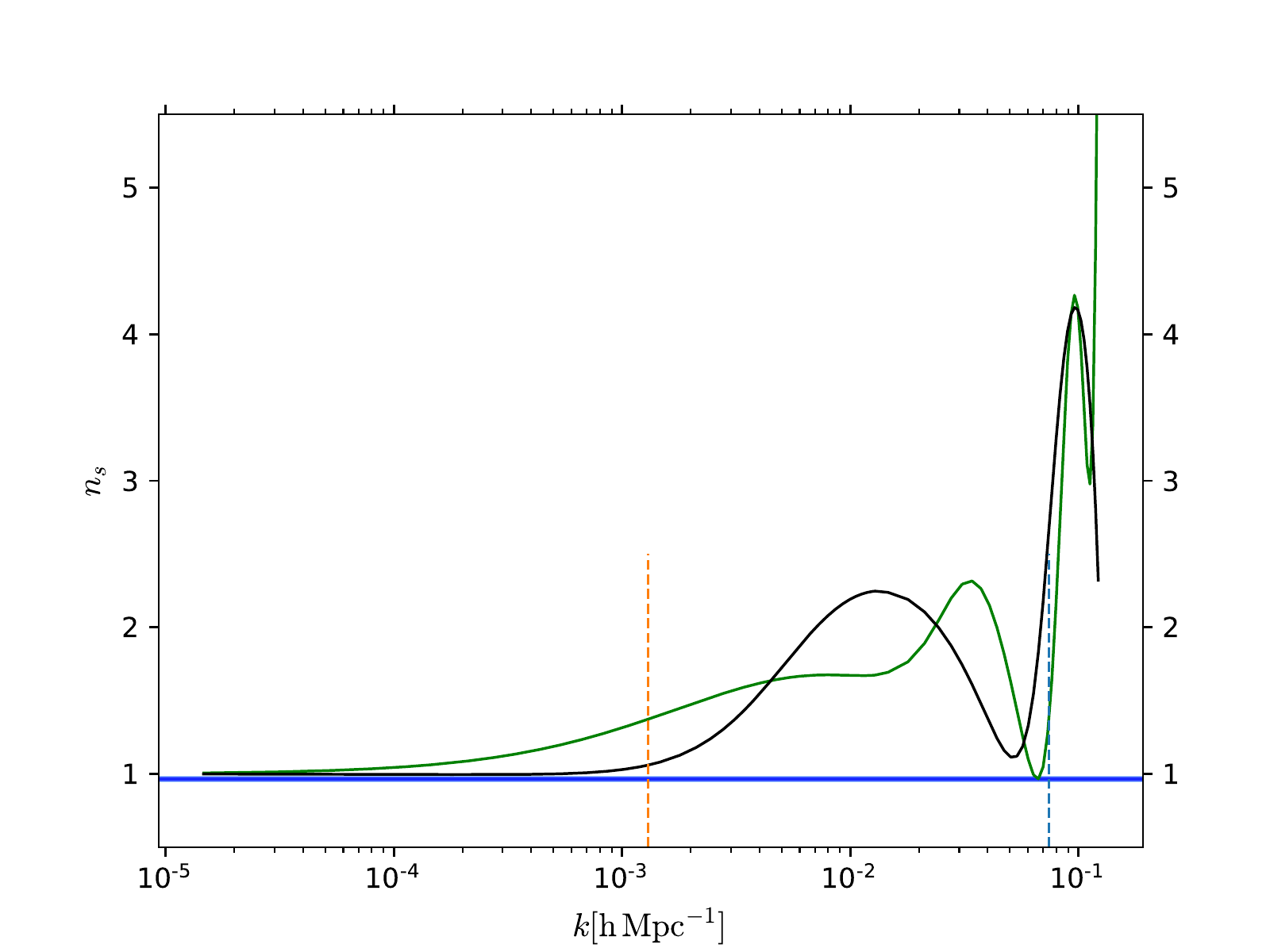}
\includegraphics[scale=0.548]{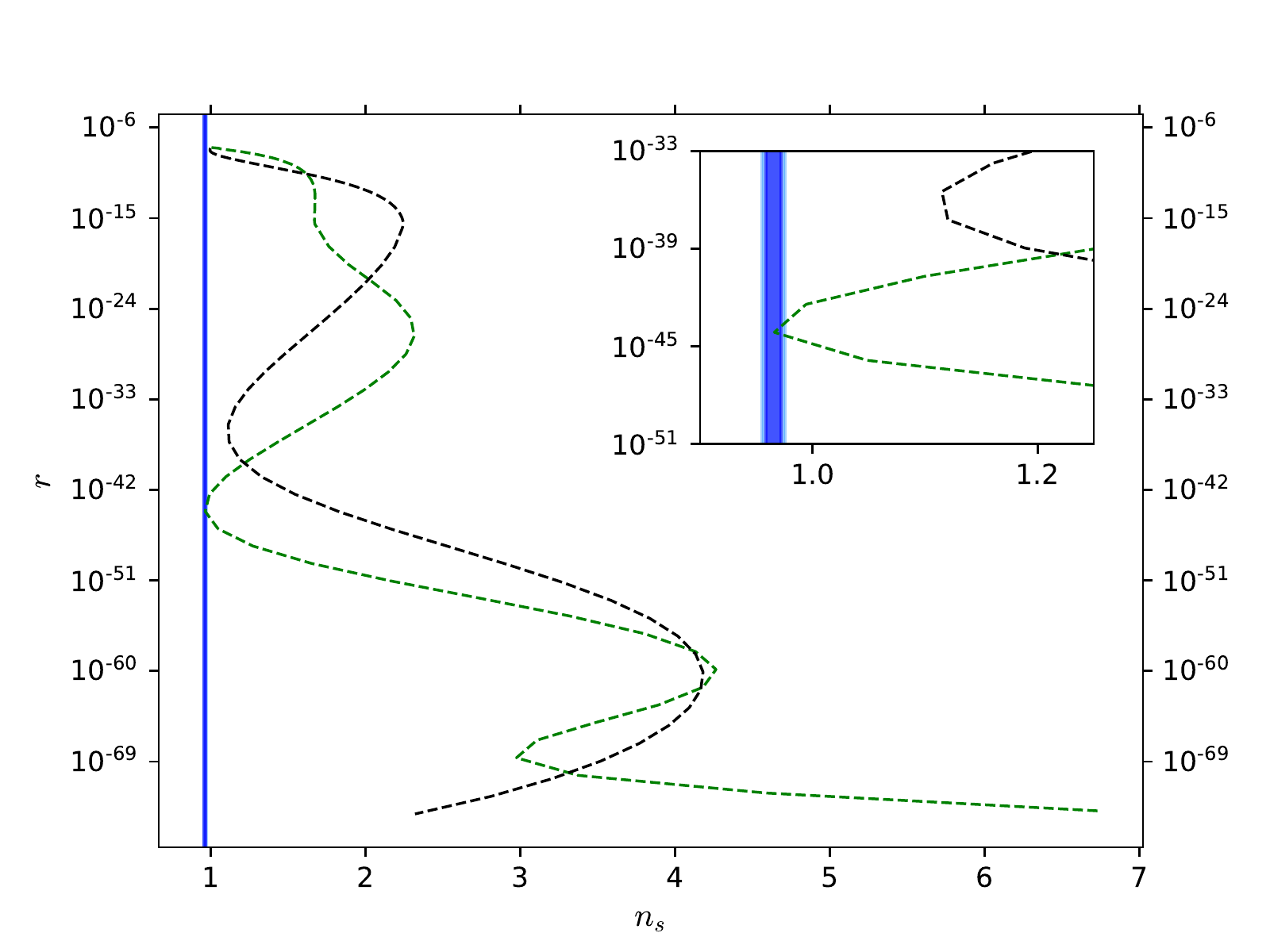} 
\caption{Left plot: evolution of the scalar spectral index $n_\mathrm{s}$ with respect to $k$. Right plot: evolution of $n_\mathrm{s}$ versus the tensor-to-scalar ratio $r$. Both panels are described by step (green) and quadratic chaotic (black) potentials for 71.0 e-folds of inflation. The orange vertical line corresponds to $k_{0}\simeq 1.3\times 10^{-3}\,h\,\rm Mpc^{-1}$, and the blue dashed line corresponds to the CMB observation by the Planck Legacy pivot value $k_{\star}\simeq 0.0743 \,h \,\rm Mpc^{-1}$ ($k_{\star}\simeq 0.05 \rm \,Mpc^{-1}$; $h=H_{0}/100 = 0.6732$, see \cite{akrami:2018b, Planck:2018jri}). The blue contours correspond to the $68\%$ and $95\%$ C.L. results from Planck 2018 TT,TE,EE+lowE+lensing data \cite{akrami:2018b, Planck:2018jri}. Note that the green line barely touches the Planck contours around $k_{\star}$.}\label{fig:WI_phys_units_ns_k}
\end{figure}

We plot the evolution of the scalar spectral index $n_\mathrm{s}$ with respect to $k$, and $n_\mathrm{s}$ versus the tensor-to-scalar ratio $r$ (see Fig.~\ref{fig:WI_phys_units_ns_k}). First, $n_\mathrm{s}$ was computed first by interpolating our numerical results, that is, using scipy.interpolate.splrep~\cite{2020SciPy-NMeth}, and then we computed the numerical derivative via Python~\cite{10.5555}: numpy.diff~\cite{harris2020array}. This time, the spectral index $n_\mathrm{s}$ is clearly blue-tilted, at smaller scales, and the tensor-to-scalar ratio $r$ becomes too low. Both results are a direct consequence of WI~\cite{Bastero-Gil:2009sdq,Bastero-Gil:2018yen,Bastero-Gil:2018uep}. However, since $P_{\mathcal{R}}$ starts oscillating around $k/k_{0}\sim 40$, $n_\mathrm{s}$ fluctuates as well, hence it can turn from blue-tilted towards red-tilted. Also, the result from the step potential (green line) skims the Planck contours.   

This time in the range $ 0.002 \lesssim k[\rm Mpc^{-1}]\lesssim 0.0035$ ($2.29 \lesssim k/k_{0} \lesssim 4.0$) $P_{\mathcal{R}}$ does not present any oscillation, but a growth; however, it oscillates around $k/k_{0}\sim 40$ ($k \simeq 0.052 \,h \,\rm Mpc^{-1}$), which is closer to the Planck pivot value $k_{\star}\simeq 0.0743 \,h \,\rm Mpc^{-1}$ ($k_{\star}\simeq 0.05 \rm \,Mpc^{-1}$) (see Fig.~\ref{fig:WI_phys_units_ns_k}).


\newpage

%
%
\section{Comparison between CI and WI}
\begin{figure}[htbp] 
\includegraphics[scale=1.0]{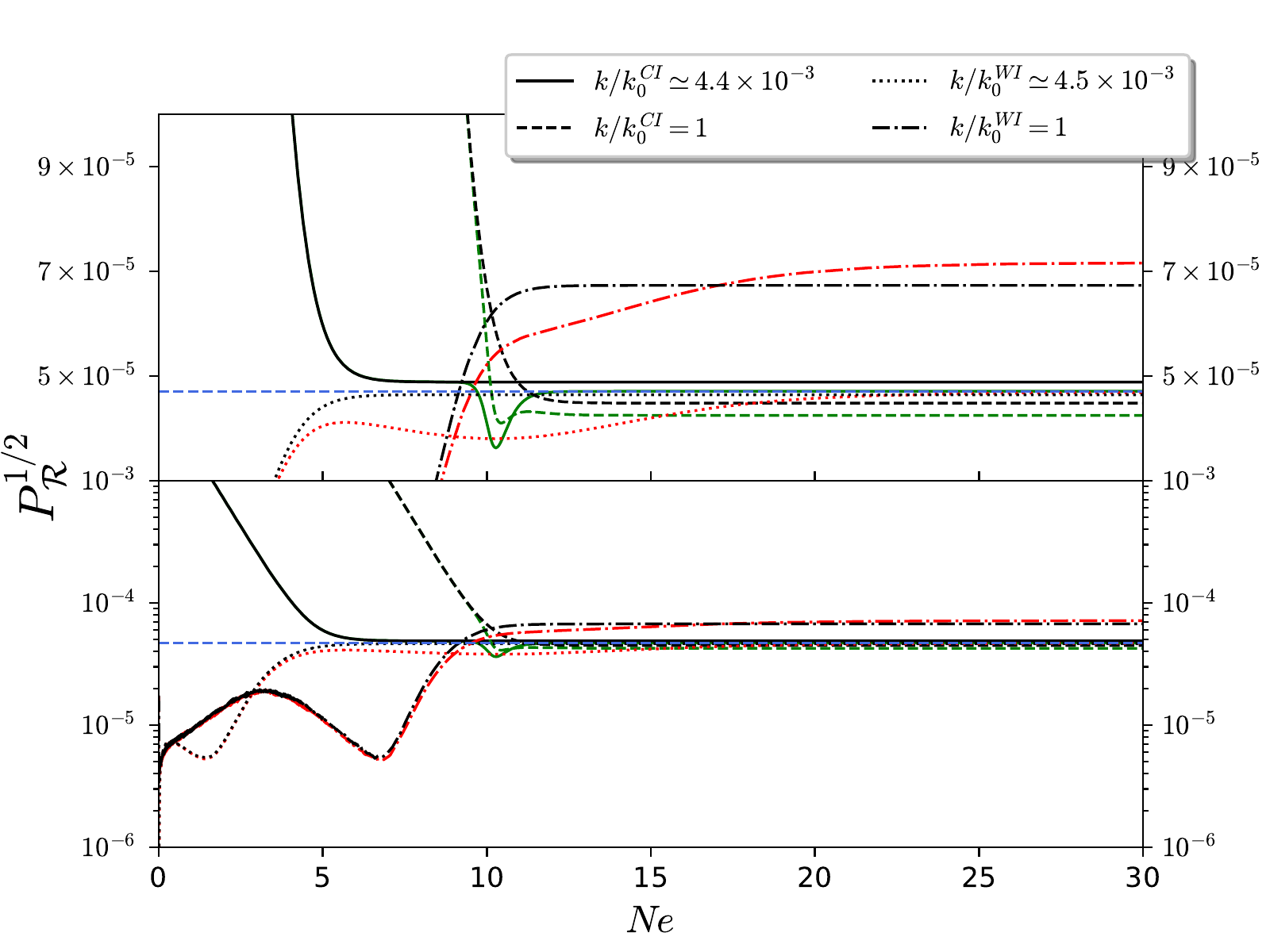}
\caption{Comparison of WI and CI behaviours of the square root of the curvature power spectra $P_{\mathcal{R}}^{1/2}$ with respect to the number of e-folds $N_{e}$, described by step (green and red) and quadratic chaotic (black) potentials for $70$ e-folds of inflation. We show the outcome for two different wave numbers $k/k_{0} \simeq 1, 4\times 10^{-3}$, and for both cold (green) and warm (red) inflation. Solid and dotted lines correspond to $N_{e}^{*}\simeq 4.6$, whilst dashed and dash-dotted lines correspond to $N_{e}^{*}\simeq 10$. The blue dashed line corresponds to the CMB observations by the Planck Legacy value $P_{\mathcal{R}}^{1/2}\simeq 4.7\times 10^{-5}$ \cite{akrami:2018b}.}\label{fig:WI_CI_PR}
\end{figure}

This small section is dedicated to studying similarities and differences between both inflationary paradigms. One key aspect of this research was to contrast the features of an inflationary potential between both cold and warm scenarios. Hence, Fig.~\ref{fig:WI_CI_PR} shows similarities and differences among various examples. At the same time, CI and WI exhibit an unequivocal step close to $\phi/\phi_{step}\simeq 1 \, (N_{e}\simeq 10)$, and several e-folds after $P_{\mathcal{R}}^{1/2}$ becomes constant. However, a major distinction between both cases is the evident growth from $k/k_{0}\sim 10^{-3}$ to $k/k_{0}\sim 1$, although preliminary, of the warm $P_{\mathcal{R}}^{1/2}$ (red lines), opposite to the CI instance (green lines).  

\begin{figure}[htbp] 
\includegraphics[scale=1.0]{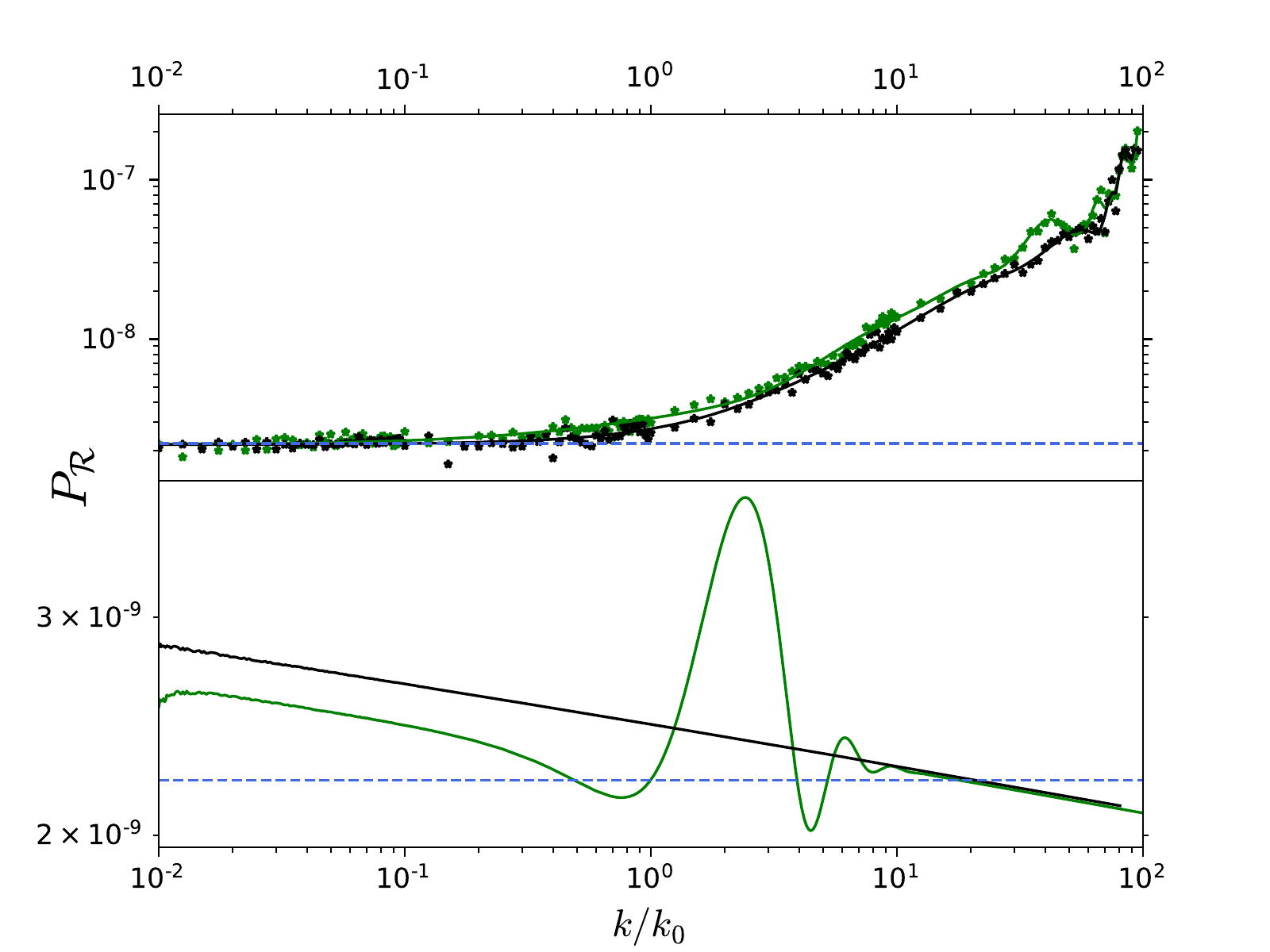}
\caption{Comparison of WI (top) and CI (bottom) behaviours of the curvature power spectra $P_{\mathcal{R}}$ with respect to the ratio $k/k_{0}$, described by step (green) and quadratic chaotic (black) potentials for $70$ e-folds of inflation. In the top panel, filled stars are the numerical values; while solid lines are the result of interpolation using a Python algorithm for curve fitting with spline functions: scipy.interpolate.splrep~\cite{2020SciPy-NMeth}. The blue dashed line corresponds to the CMB observations by the Planck Legacy value $P_{\mathcal{R}}\simeq 2.22\times 10^{-9}$ \cite{akrami:2018b}. Note that WI displaces the oscillatory nature of the featured potential from $k/k_{0}\sim 40$ onwards, and the chaotic scenario wiggles too; however, further studies beyond $k/k_{0}> 100$ are required to corroborate that a chaotic potential also fluctuates at smaller scales.}\label{fig:WI_CI_PR_k_k0}
\end{figure}

Fig.~\ref{fig:WI_CI_PR_k_k0} shows one last comparison of the evolution of $P_{\mathcal{R}}$ with respect to $k/k_{0}$. Immediately, the most notorious distinction between both schemes is the growth of $P_{\mathcal{R}}$ within the WI scenario. This upshot shows that the amplitude of the primordial spectrum can be much larger than the CMB value $P_{\mathcal{R}}\simeq 2.22\times 10^{-9}$ \cite{akrami:2018b} at small scales. Moreover, the oscillatory behaviour described by CI appears only around $k/k_{0}\sim 1$, while in WI $P_{\mathcal{R}}$ fluctuates at $k/k_{0}\sim 40$, and, in fact, both potentials change. However, further research beyond $k/k_{0}> 100$ is required to corroborate that a chaotic potential also fluctuates at smaller scales; and therefore, WI does screen the step of the featured potential.

%

%
%
\section{Final remarks}
The main purpose of this paper was to thoroughly analyse how an inflationary scenario from a chaotic potential with a step affects the relevant observational parameters. We work from two fronts: CI and WI. 

In general terms, in the CI case, the step indeed produces oscillations in the primordial power spectrum, particularly around $k_{0}$ (see Fig.~\ref{fig:CI_PR_k}), which belongs to the small-scale perturbation sector. Then $P_{\mathcal{R}}$ evolves equally for both chaotic and step potentials. Although many of our results are very similar to \cite{adams:2001,hunt:2004,covi:2006,hamann:2007,chen:2007,hunt:2007,hamann:2010,bartolo:2013,cadavid:2015,cadavid:2017}, and in particular to \cite{cadavid:2017}; we have included the analysis of $n_\mathrm{s}$ and $r$. Indeed, we found that more than one region of ($n_\mathrm{s},r$) lies within the observable contours \cite{akrami:2018b, Planck:2018jri}. And in particular, from Fig.~\ref{fig:CI_r_k} one can compute the minimum of $r\simeq 0.084$ that corresponds to $k/k_{0}\simeq 2.43$. However, the choice of the pivot scale determines the observational bounds on $n_\mathrm{s}$ and $r$, hence, according to the range $k = 0.002 \sim 0.0035 \rm Mpc^{-1}$ our results do not contain any hints of oscillations of $P_{\mathcal{R}}$ (see Appendix~\ref{appendix_a}).  

On the other hand, in the WI scenario, a straightforward result is the growth of the amplitude of the power spectrum before horizon crossing; this is due to the stochastic source $\Xi_{k}\equiv \Xi(\mathbf{k},t)$. Furthermore, the feature of the potential indeed induces perceptible changes in $P_{\mathcal{R}}$ around the step ($N_{e}\simeq 10$) (see Fig.~\ref{fig:WI_PR}). Moreover, one can observe that the step enhances the evolution of $P_{\mathcal{R}}^{1/2}$ when $k/k_{0} \simeq 1$, where for a short period of time it slightly disconnects from the chaotic quadratic path, but when both amplitudes become constant, their values are essentially equal. Also, this time the spectral index $n_\mathrm{s}$ is clearly blue-tilted at smaller scales, and the tensor-to-scalar ratio $r$ becomes too low. Both results are a direct consequence of WI~\cite{Bastero-Gil:2009sdq,Bastero-Gil:2018yen,Bastero-Gil:2018uep}. Nonetheless, $n_\mathrm{s}$ can change from blue-tilted towards red-tilted, since $P_{\mathcal{R}}$ starts to oscillate around $k/k_{0}\sim 40$ ($k \simeq 0.052 \,h \,\rm Mpc^{-1}$), which is closer to the Planck pivot value $k_{\star}\simeq 0.0743 \,h \,\rm Mpc^{-1}$ ($k_{\star}\simeq 0.05 \rm \,Mpc^{-1}$). Indeed, the result from the step potential (green line of Fig.~\ref{fig:WI_phys_units_ns_k}) skims the Planck contours.   

One remarkable upshot in both scenarios is that certain fluctuation scales are not longer ``freeze in'' on super-horizon scales. In CI this effect is due to a featured background~\cite{Vallejo-Pena:2019lfo, Gordon:2000hv}; and in WI $\mathcal{R}$ continues to evolve a few e-folds after horizon crossing through a combined effect from the background (as in CI)~\cite{Vallejo-Pena:2019lfo, Gordon:2000hv} and the entropy perturbations~\cite{DeOliveira:2001he,Gordon:2000hv,Wands:2000dp}. Notably, WI boosts the growth of the amplitude of the primordial spectrum, whilst moves the oscillatory behaviour from $k/k_{0}\sim 40$ onwards (see Fig.~\ref{fig:WI_CI_PR_k_k0}); although, not only the step potential wiggles but also the chaotic. Further research beyond $k/k_{0}> 100$ is required to corroborate that a chaotic potential also fluctuates at smaller scales and, therefore, WI might not screen the step of the featured potential. 

We have confirmed that perturbations are amplified in WI, and recently this framework has been proposed as a plausible mechanism that could lead to the formation of Primordial Black Holes on re-entry, provided the amplitude reaches a critical value $P_{\mathcal{R}}\sim 10^{-2}$~\cite{Arya:2019wck,Bastero-Gil:2021fac}. This opens up new paths that we can explore. 

%
%
\appendix
\section{A CI example with $k_0 = 0.0014 \,\rm Mpc^{-1}$ $(0.002 \,h \,\rm Mpc^{-1})$}\label{appendix_a}
\begin{figure}[htbp] 
\includegraphics[scale=1.0]{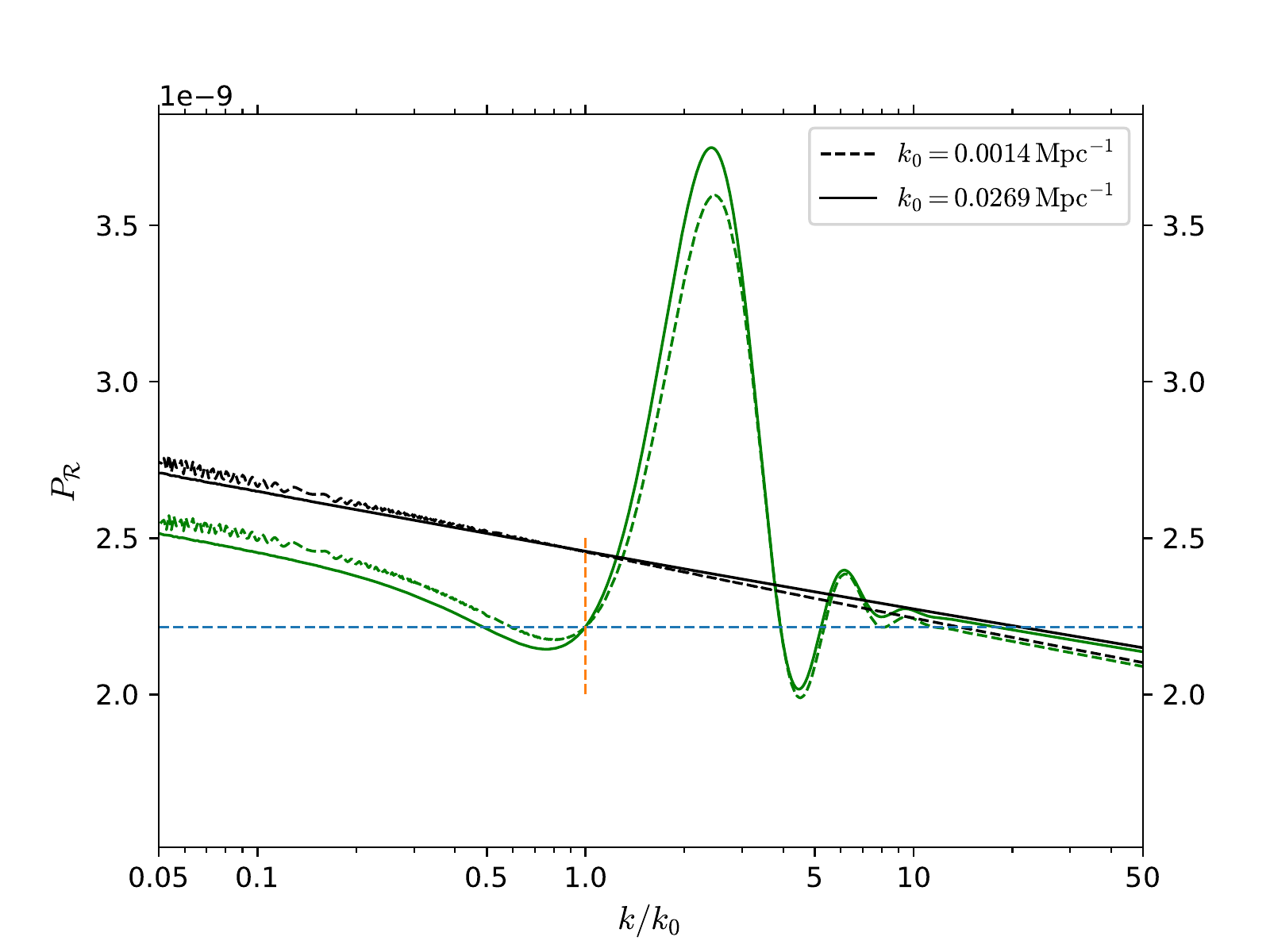}
\caption{Behaviour of the curvature power spectra $P_{\mathcal{R}}$ with respect to the ratio $k/k_{0}$, where the solid lines correspond to $k_{0}= 0.0269 \rm Mpc^{-1}$ $(0.04 \,h \,\rm Mpc^{-1})$, whilst the dashed lines correspond to $k_{0}= 0.0014 \rm Mpc^{-1}$ $(0.002 \,h \,\rm Mpc^{-1})$. Both examples are described by step (green) and quadratic chaotic (black) potentials for $71.0$ (solid) and $60.13$ (dashed) e-folds of inflation. The orange vertical line corresponds to $k/k_{0}=1$ and $P_{\mathcal{R}}\simeq 2.22\times 10^{-9}$. On the other hand, the blue dashed line corresponds to the CMB observations by the Planck Legacy value $P_{\mathcal{R}}\simeq 2.22\times 10^{-9}$ \cite{akrami:2018b}.}\label{fig:appendix_CI_PR_k_k0}
\end{figure}

\begin{figure}[htbp] 
\includegraphics[scale=1.0]{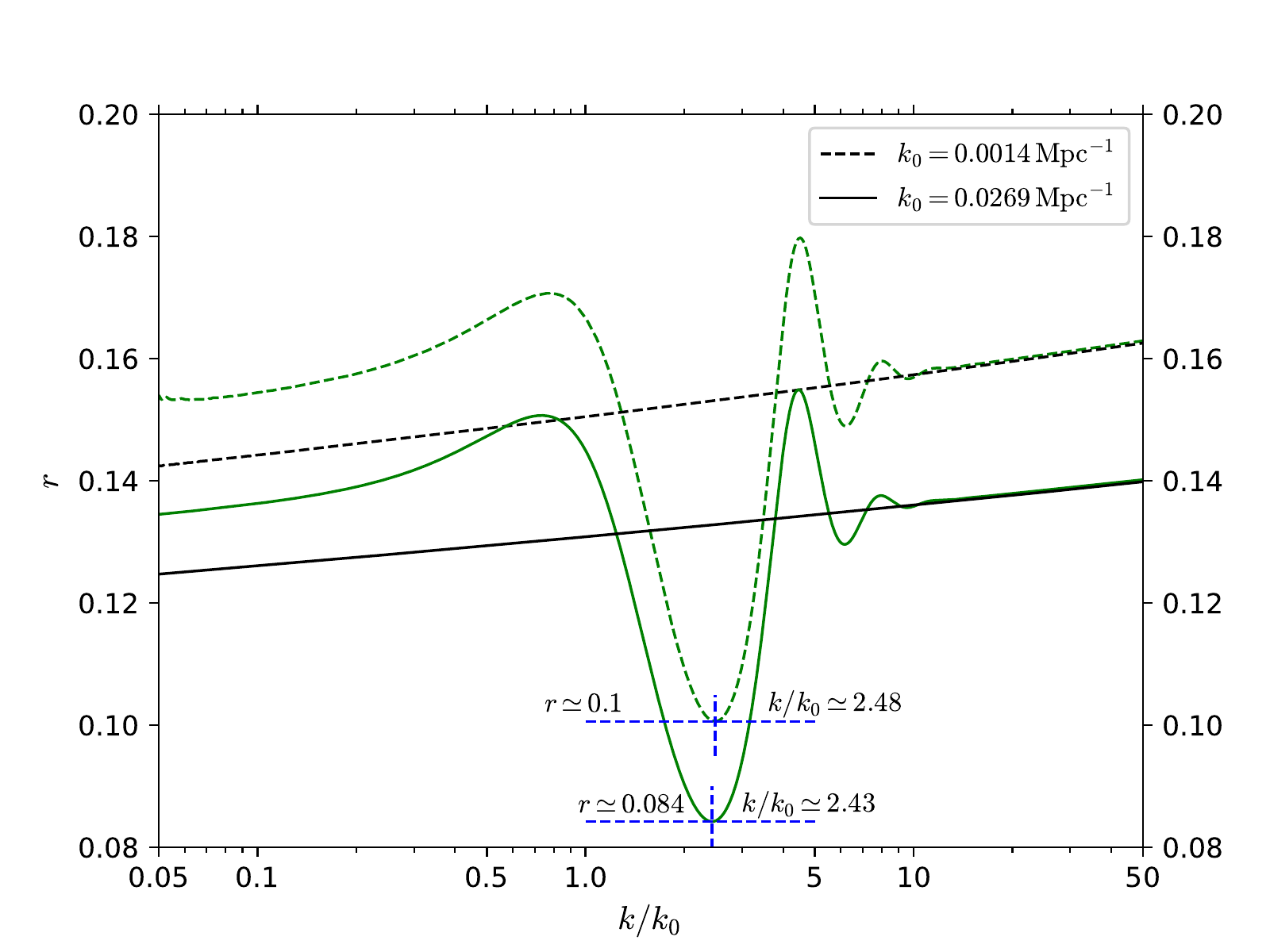}
\caption{Behaviour of the tensor-to-scalar ratio $r$ with respect to $k/k_{0}$, where solid lines correspond to $k_{0}= 0.0269 \rm Mpc^{-1}$ $(0.04 \,h \,\rm Mpc^{-1})$, whilst dashed lines correspond to $k_{0}= 0.0014 \rm Mpc^{-1}$ $(0.002 \,h \,\rm Mpc^{-1})$. Both examples are described by step (green) and quadratic chaotic (black) potentials for $71.0$ (solid) and $60.13$ (dashed) e-folds of inflation. Note that the minimum value, with $k_{0}= 0.0269 \rm Mpc^{-1}$, $r\simeq 0.084$ corresponds to $k/k_{0}\simeq 2.43$, and with $k_{0}= 0.0014 \rm Mpc^{-1}$ we have $r\simeq 0.1$ at $k/k_{0}\simeq 2.48$. In both cases $r$ lies within the Planck contours \cite{akrami:2018b}.}\label{fig:appendix_CI_r_k_k0}
\end{figure}

Due to the insightful comments of the anonymous reviewer, we realised that to examine the hints of a dip and a bump in the spectrum of primordial perturbations, by means of the features of the potential, the range of scales $k = 0.002 \rm Mpc^{-1}$ and $k = 0.0035 \rm Mpc^{-1}$ are used~\cite{DiValentino:2016ikp,Benetti:2016tvm,GallegoCadavid:2016wcz}. Hence, we decide to examine a particular example with $k_0 = 0.0014 \,\rm Mpc^{-1}$ $(0.002 \,h \,\rm Mpc^{-1})$, and we scan the range $0.05 \leq k/k_0 \leq 50$. We take $m=7.255\times 10^{-6}\, M_{Pl}$, $\phi_{\textnormal{step}}=14.43 M_{Pl}$, $c=0.0012$, $d=0.04  \, M_{Pl}$, and the initial value of the inflaton $\phi_{0}=15.45 \, M_{Pl}$. Inflation lasts $N_{e}=60.13$, and the step occurs at $\phi=\phi_{\textnormal{step}}$ around $N_{e}\simeq 7.62$, so from the step $\Delta N_{e}=52.51$. The general behaviours of $P_{\mathcal{R}}$ (Fig.~\ref{fig:appendix_CI_PR_k_k0}) and $r$ (Fig.~\ref{fig:appendix_CI_r_k_k0}) are qualitatively the same as the results with $k_{0}= 0.0269 \rm Mpc^{-1}$ $(0.04 \,h \,\rm Mpc^{-1})$, and consequently $n_\mathrm{s}$ also presents the same outcome. However, as the reviewer remarked, the resulting observational values are, in fact, the consequence of the initial values of the parameters. So, for instance, the minimum value of $r$ is now $r\simeq 0.1$ at $k/k_{0}\simeq 2.48$ (see Fig.~\ref{fig:appendix_CI_r_k_k0}).

%

\begin{acknowledgements}
The authors thank the anonymous reviewer for helping us to improve our paper. R.H.J is supported by CONACYT Estancias Posdoctorales por M\'{e}xico, Modalidad 1: Estancia Posdoctoral Acad\'{e}mica.
\end{acknowledgements}

%
%
\bibliographystyle{unsrt}
\bibliography{main4}


\end{document}